\documentclass[10pt,journal,twocolumn]{IEEEtran}
\IEEEoverridecommandlockouts
\usepackage{cite}
\usepackage{amsmath,amssymb,amsfonts,booktabs}
\usepackage{algorithmic}
\usepackage{graphicx}
\usepackage{textcomp}
\usepackage{xcolor}
\usepackage{enumerate}
\usepackage{epstopdf}
\usepackage{subcaption}
\usepackage{amsmath,amsfonts}
\usepackage{amsthm}
\usepackage{amssymb}
\usepackage{easyReview}
\usepackage{algorithmic}
\usepackage{algorithm}
\usepackage{array}
\usepackage[caption=false,font=normalsize,labelfont=sf,textfont=sf]{subfig}
\usepackage{easyReview}
\usepackage{textcomp}
\usepackage{stfloats}
\usepackage{url}
\usepackage{bm}
\usepackage{makecell}
\usepackage{verbatim}
\usepackage{graphicx}
\usepackage{cite}
\usepackage{caption}
\usepackage{booktabs}
\usepackage{amsthm}
\newtheorem{theorem}{Theorem}
\newtheorem{lemma}{Lemma}
\newtheorem{remark}{Remark}
\newtheorem{proposition}{Proposition}

\usepackage{array}
\definecolor{mygray}{RGB}{240,240,240}

\usepackage{algorithm}
\usepackage{algorithmic}
\allowdisplaybreaks[4]

\def\BibTeX{{\rm B\kern-.05em{\sc i\kern-.025em b}\kern-.08em
		T\kern-.1667em\lower.7ex\hbox{E}\kern-.125emX}}
\begin{document}

\title{Movable-Antenna Empowered AAV-Enabled Data Collection over Low-Altitude Wireless Networks}

\author{
        Xuhui Zhang,
        Wenchao Liu,
        Jinke Ren,
        Chunjie Wang,
        Huijun Xing,
        Yanyan Shen,
        and Shuguang Cui

\thanks{
Xuhui Zhang and Jinke Ren are with the Shenzhen Future Network of Intelligence Institute, the School of Science and Engineering, and the Guangdong Provincial Key Laboratory of Future Networks of Intelligence, The Chinese University of Hong Kong, Shenzhen, Guangdong 518172, China (e-mail: xu.hui.zhang@foxmail.com; jinkeren@cuhk.edu.cn).
}

\thanks{
Wenchao Liu is with the School of Automation and Intelligent Manufacturing, Southern University of Science and Technology, Shenzhen 518055, China (e-mail: wc.liu@foxmail.com).
}

\thanks{
Huijun Xing is with the Department of Electrical and Electronic Engineering, Imperial College London, London SW7 2AZ, The United Kingdom (e-mail: huijunxing@link.cuhk.edu.cn).
}

\thanks{
Chunjie Wang and Yanyan Shen are with Shenzhen Institutes of Advanced Technology, Chinese Academy of Sciences, Guangdong 518055, China (e-mail: cj.wang@siat.ac.cn; yy.shen@siat.ac.cn).
}

\thanks{
Shuguang Cui is with the School of Science and Engineering, the Shenzhen Future Network of Intelligence Institute, and the Guangdong Provincial Key Laboratory of Future Networks of Intelligence, The Chinese University of Hong Kong, Shenzhen, Guangdong 518172, China (e-mail: shuguangcui@cuhk.edu.cn).
}

}

\maketitle

\begin{abstract}
    Movable-antennas (MAs) are revolutionizing spatial signal processing by providing flexible beamforming in next-generation wireless systems.
    This paper investigates an MA-empowered autonomous aerial vehicle (AAV) system in low-altitude wireless networks (LAWNs) for uplink data collection from ground users.
    We aim to maximize the sum achievable rate by jointly optimizing the AAV trajectory, receive beamforming, and MA positions. An efficient alternating optimization (AO) algorithm that incorporates successive convex approximation, weighted minimum mean square error, and particle swarm optimization is developed. The analysis of the computational complexity and convergence features is provided.
    Extensive simulations demonstrate superior performance in terms of the sum achievable rate and the service reliability comparing to several benchmark schemes.
    These results demonstrate the distinctive advantages of the proposed scheme: enhanced spectral efficiency via adaptive beam-user alignment and improved collection reliability through spatial interference management, highlighting the implementation potential of the MA-empowered LAWNs.

\end{abstract}


\section{Introduction}
\IEEEPARstart {T}{he} burgeoning low-altitude wireless networks (LAWNs) have emerged as a pivotal enabler for supporting low-altitude economy (LAE) applications, unlocking transformative potential in applications such as emergency logistics, environmental monitoring, and infrastructure inspection \cite{10879807, 10693833}.
Autonomous aerial vehicles (AAVs), also known as
unmanned aerial vehicles (UAVs), as the key aerial devices in the LAWNs, provide dynamic coverage extension, on-demand service deployment, and adaptive resource allocation, thanks to their high mobility and maneuverability \cite{9456851}.
These capabilities allow AAVs to bridge connectivity gaps in remote areas and enhance situational awareness for time-critical missions \cite{8918497, 10606316}.
However, the escalating demand for spectrum-intensive tasks, such as high-resolution sensing, latency-sensitive monitoring, and ultra-reliable communications, exacerbates the scarcity of communication resources.
Existing single-antenna AAV systems, constrained by limited spatial multiplexing and interference management, fail to meet the throughput requirements of next-generation LAE systems.

Multi-antenna technologies, particularly antenna beamforming, offer a paradigm shift in the AAV-enabled LAWNs by enabling spatial multiplexing, interference suppression, and enhanced spectral efficiency \cite{7894280}. AAVs equipped with multiple antennas can dynamically adjust radiation patterns to focus energy toward target users, significantly improving signal-to-interference-plus-noise ratio (SINR) and throughput \cite{8030501}. This capability is critical for supporting dense Internet of things (IoT) deployments and mobile ground users in the LAE scenarios \cite{9461747}.
However, in the AAV-enabled LAWNs, the inherent randomness of user mobility and channel dynamics introduces rapid channel variations and spatially non-stationary interference, resulting in inefficient communication resource utilization.
Existing beamforming techniques, designed primarily for static or predictable environments, struggle to maintain precise beam alignment under such dynamic conditions, leading to suboptimal resource utilization and degraded quality of service (QoS) \cite{10972180}.

Movable-antennas (MAs), also referred to fluid antennas \cite{zhu2024historical}, further revolutionize the AAV-enabled communication systems by enabling dynamic beam reconfiguration with spatial antenna gains \cite{10286328, liu2025movable}.
Unlike fixed-position antennas (FPAs), the MAs allow real-time adjustment of radiation patterns to adapt to user mobility and channel dynamics, thereby enhancing interference mitigation and energy efficiency \cite{10318061, 11017619}.
When integrated with the AAVs, the MAs unlock synergistic benefits, namely, the aerial mobility of the AAVs complements the beam agility of the MAs, enabling precise alignment with distributed ground users and IoT devices \cite{bai2024movable}.
This hybrid mobility ensures robust connectivity in rapidly changing environments while maintaining low latency and high throughput.
Moreover, the AAVs equipped with MAs can exploit spatial-temporal correlations in channel states, optimizing spectrum utilization beyond conventional multi-antenna architectures.

Inspired by the beamforming gains of the MAs, we investigate an AAV-enabled LAWN empowered by the MAs.
Meanwhile, the rapid evolution of generative artificial intelligence (GAI) has spurred unprecedented demands for centralized data processing, as massive distributed users require reliable uplink transmission to support latency-sensitive applications. Unlike prior studies focusing on the downlink transmission scenarios, we address the critical yet underexplored challenge of uplink transmission and aim to maximize the data collection rate of ground users in the AAV-enabled LAWN, ensuring efficient spectrum utilization under dynamic channel conditions. The key contributions of this paper are summarized as follows:
\begin{itemize}
    \item We study a novel MA-empowered uplink communication over the LAWNs, where an AAV is dispatched to collect data from ground users. We formulate an optimization problem to maximize the sum achievable rate of all ground users by jointly optimizing the AAV trajectory, the receive beamforming, the user transmit power, and the antenna positions of the MAs.
    \item To solve the non-convex problem, we decouple the problem into three sub-problems. Under the alternating optimization (AO) manner, we develop an efficient scheme to optimize the three sub-problems iteratively.
    The computational complexity and convergence behavior of the proposed scheme are also analyzed.
    \item We conduct extensive simulations to compare the performance of the proposed scheme with several benchmark schemes. The results demonstrate consistent superiority in sum achievable rate across diverse metrics, validating the efficacy of our proposed scheme. 
\end{itemize}

\textit{Organizations:}
The remainder of this paper is organized as follows.
Section II surveys the related works. Section III presents the MA-empowered AAV data collection system for the LAWNs and formulates the sum-rate maximization problem. The AO-based algorithm is developed in Section IV, with complexity and convergence analyses. Performance evaluation is conducted in Section V, followed by concluding remarks in Section VI.

\textit{Notations:}
The following notations are adopted in this work.
Let $\mathrm{j}$ satisfy $\mathrm{j}^2 = -1$ as the imaginary unit.
For complex $z$, $\Re{z}$ denotes its real part.
$\mathbb{C}^{M\times N}$ represents the space of $M \times N$ complex matrices.
For matrix $\mathbf{G}$, $\mathbf{G}^{\mathsf{H}}$ and $\mathbf{G}^{\mathsf{T}}$ denote conjugate transpose and transpose, respectively.
For vector $\mathbf{w}$, $|\mathbf{w}|$ is the Euclidean norm, and $\mathrm{diag}(\mathbf{w})$ constructs a diagonal matrix from $\mathbf{w}$.
$\mathcal{CN}(\mu, \sigma^2)$ indicates a circularly symmetric complex Gaussian distribution with mean $\mu$ and variance $\sigma^2$.

\section{Related Works}
In this section, we review the existing literature across three interconnected research areas: the AAV-enabled LAWNs, the AAV-enabled data collection systems, and the MA-empowered AAV communications. This structured review establishes the technological landscape motivating our integrated approach.
To facilitate reading, a comparison between our work and related works is presented in Table \ref{table1}.

\subsection{AAV-Enabled LAWNs}
AAVs have emerged as pivotal elements in the next-generation wireless networks, primarily due to their unique capabilities in on-demand deployment and line-of-sight (LoS) connectivity.
Early works \cite{8807386, 8676325} established AAVs as aerial base stations (BSs) for coverage extension in terrestrial networks, demonstrating significant gains in edge-user capacity during temporary events.
Meanwhile, several works \cite{8626132, 9206550} investigated the AAV-supported relay systems, overcoming  terrestrial propagation barriers by establishing dynamic LoS links, thereby enhancing coverage reliability in obstructed environments while minimizing multi-hop latency.
Notably, by exploiting the maneuverability inherent to AAVs, optimized trajectory control can provide dynamic adaptation to spatial channel variations to further enhance signal receiving, and energy-efficient coverage provisioning through altitude-dependent path loss minimization \cite{9273074, 10087216}.
Several works also studied the AAV-enabled LAWNs with integrated sensing and communication (ISAC) \cite{9916163, 10233771}, localization \cite{10146439}, non-terrestrial networks \cite{9822386, 9904508, 10233456}, and secure communications \cite{10271264, 9903838},
to provide heterogeneous services for different users.
While substantial efforts have focused on downlink-oriented applications, e.g., coverage extension and content relay delivery, the burgeoning data explosion from ground users and IoT devices necessitates rethinking AAV deployment for uplink-dominated applications, particularly in latency-sensitive industrial monitoring and large-scale sensing scenarios.

\subsection{AAV-Enabled Data Collection}
Building upon uplink-centric transmission requirements, the AAV-enabled data collection has emerged as a key solution for next-generation wireless networks \cite{9604506, 10539623}.
Its aerial mobility enables strong LoS-link-enabled data harvesting from ground users with lower energy consumption than the static BSs \cite{9779853, 10980172}.
Supported by wireless charging stations, the AAV can establish seamless coverage for data collection services to enhance system throughput \cite{9321340}.
Accordingly, the AAV-enabled mobile edge computing (MEC) systems fully utilize the uplink transmission capability for data offloading to the MEC server co-located with the AAVs, or relayed by them \cite{8956055, 10100681}.
Meanwhile, AAV-enabled over-the-air computation enables simultaneous analog aggregation of collected data through coherent signal superposition, achieving lower latency and energy consumption than conventional digital data aggregation \cite{9985993, 10946250}.
Besides, through uplink transmission, federated learning can be trained by the AAV-enabled systems, where the AAVs collect the local trained models from ground users, and then aggregate the global model \cite{10220154, 10818523, 10972043}.
While AAV-enabled wireless transmission enhances communication rates to some extent, many existing works adopt fixed single-antenna AAVs. This approach inherently lacks multi-antenna beamforming capabilities for throughput improvement, and the static antenna arrangement fundamentally constrains communication performance.

\subsection{MA-Empowered AAV Communications}
Embedding the MAs onto AAV platforms creates a mutually augmented architecture, i.e., the AAV mobility expands positioning flexibility, while the MA reconfigurability enables user-oriented flexible beamforming, jointly unlocking precision beam alignment unattainable in conventional systems.
Notably, emerging works demonstrated that the MA-empowered AAV systems achieve substantial improvements in both communication throughput and interference mitigation capabilities.
Specifically, in \cite{10654366}, an MA-empowered AAV-enabled downlink transmission system was considered, and the achievable data rate was maximized by optimizing the antenna position, beamforming, and AAV trajectory.
Moreover, a six-dimensional MA array was adopted in \cite{10918750} to enhance the interference mitigation for the AAV communication systems.
Consequently, we are motivated to investigate user-centric uplink transmission in AAV-enabled data collection systems, where the synergistic integration of the MA mobility, the receive beamforming, and the AAV trajectory design achieves data collection rate enhancement for the collection tasks.

\begin{table*}[t]
	\caption{Comparison of Selected Related Works}
	\centering
	\label{table1}
	\resizebox{\textwidth}{!}{
		\renewcommand{\arraystretch}{2.3} 
		\begin{tabular}{|c|c|c|c|c|c|c|c|}
			\hline
			\textbf {Ref.} & \textbf{Scenario} & \textbf{Objective} & \textbf{Optimization variables} & \textbf{AAV mobility} & \textbf{MAs} & \textbf{Data collection} \\
                \hline
			\cite{8807386} & \makecell{AAV-Enabled LAWNs} & \makecell{Maximizing the long-term rewards} & \makecell{User selection, power \\ level, and subchannel selection} & No & No & No \\
			\hline
                \cite{8676325} & \makecell{AAV-enabled LAWNs} & \makecell{Maximizing the temporal average coverage scores} & \makecell{AAV flying direction and distance} & Yes & No & No \\
			\hline
                \cite{8626132} & \makecell{AAV-enabled LAWNs} & \makecell{Maximizing the minimum average secrecy rate} & \makecell{AAV trajectory and time scheduling} & Yes & No & No \\
                \hline
                \cite{9206550} & \makecell{AAV-enabled LAWNs} & \makecell{Minimizing the total decoding error rate} & \makecell{AAV trajectory, RIS phase shifts, and latency} & Yes & No & No \\
                \hline
                \cite{9916163} & \makecell{AAV-enabled ISAC system} & \makecell{Maximizing the weighted sum-rate} & \makecell{AAV maneuver and transmit beamforming} & Yes & No & No \\
			\hline
                \cite{10146439} & \makecell{AAV-enabled LAWNs} & \makecell{Maximizing the sum communication rate} & \makecell{AAV deployment,  bandwidth and power allocation} & Yes & No & No \\
                \hline
                \cite{9904508} & \makecell{AAV-non-terrestrial networks} & \makecell{Maximizing the system utility} & \makecell{Price decision and bandwidth allocation} & No & No & No \\
                \hline
                \cite{10271264} & \makecell{AAV-enabled secure systems} & \makecell{Maximizing the achievable transmission rate} & \makecell{Transmit power and jamming power} & No & No & No \\
                \hline
                \cite{10539623} & \makecell{AAV-enabled data collection} & \makecell{Minimizing the average age of information} & \makecell{AAVs' beamforming, users' access \\ control, and trajectory planning strategies} & Yes & No & Yes \\
			\hline
                \cite{10980172} & \makecell{AAV-enabled data collection} & \makecell{Minimizing the average age of information} & \makecell{AAV trajectory, collection scheduling, and completion time} & Yes & No & Yes \\
			\hline
                \cite{8956055} & \makecell{AAV-enabled data collection} & \makecell{Minimizing the weighted sum of the service delay \\ and AAV energy consumption} & \makecell{AAV position, communication and computing \\ resource allocation, and task splitting decisions} & Yes & No & Yes \\
			\hline
                \cite{9985993} & \makecell{AAV-enabled data collection} & \makecell{Maximizing the minimum amount of tasks} & \makecell{AAV trajectory, transceiver design, \\ cluster scheduling and association} & Yes & No & Yes \\
                \hline
                \cite{10946250} & \makecell{AAV-enabled data collection} & \makecell{Minimizing the mean square error} & \makecell{AAV deployment and precoding coefficients of sensors} & Yes & No & Yes \\
                \hline
                \cite{10972043} & \makecell{AAV-enabled data collection} & \makecell{Minimizing the latency} & \makecell{AAV trajectory, bandwidth allocation, \\ computing resources, and transmit power} & Yes & No & Yes \\
                \hline
                \cite{10654366} & \makecell{MA-AAV communications} & \makecell{Maximizing the achievable data rate} & \makecell{AAV trajectory, transmit \\ beamforming, and the positions of the MA} & Yes & Yes & No \\
                \hline
                \cite{10918750} & \makecell{MA-AAV communications} & \makecell{Maximizing the SINR} & \makecell{Antenna position vector, array rotation vector, \\ receive beamforming vector, and associated BS of the AAV} & No & Yes & No \\
			\hline
			\textbf{Our work} & \makecell{\textbf{MA-empowered AAV-enabled} \\\textbf{data collection}} & \makecell{\textbf{Maximizing the sum-rate of users}} & \makecell{\textbf{AAV trajectory, receive beamforming,} \\\textbf{MA position, and user transmit power}} & \textbf{Yes} & \textbf{Yes} & \textbf{Yes} \\
			\hline
	\end{tabular}}
    \renewcommand{\arraystretch}{2.3} 
\end{table*}

\section{System Model and Problem Formulation}

\subsection{System Model}
\begin{figure}[!htbp]
	\centering
	\fbox{\includegraphics[width=0.88\linewidth]{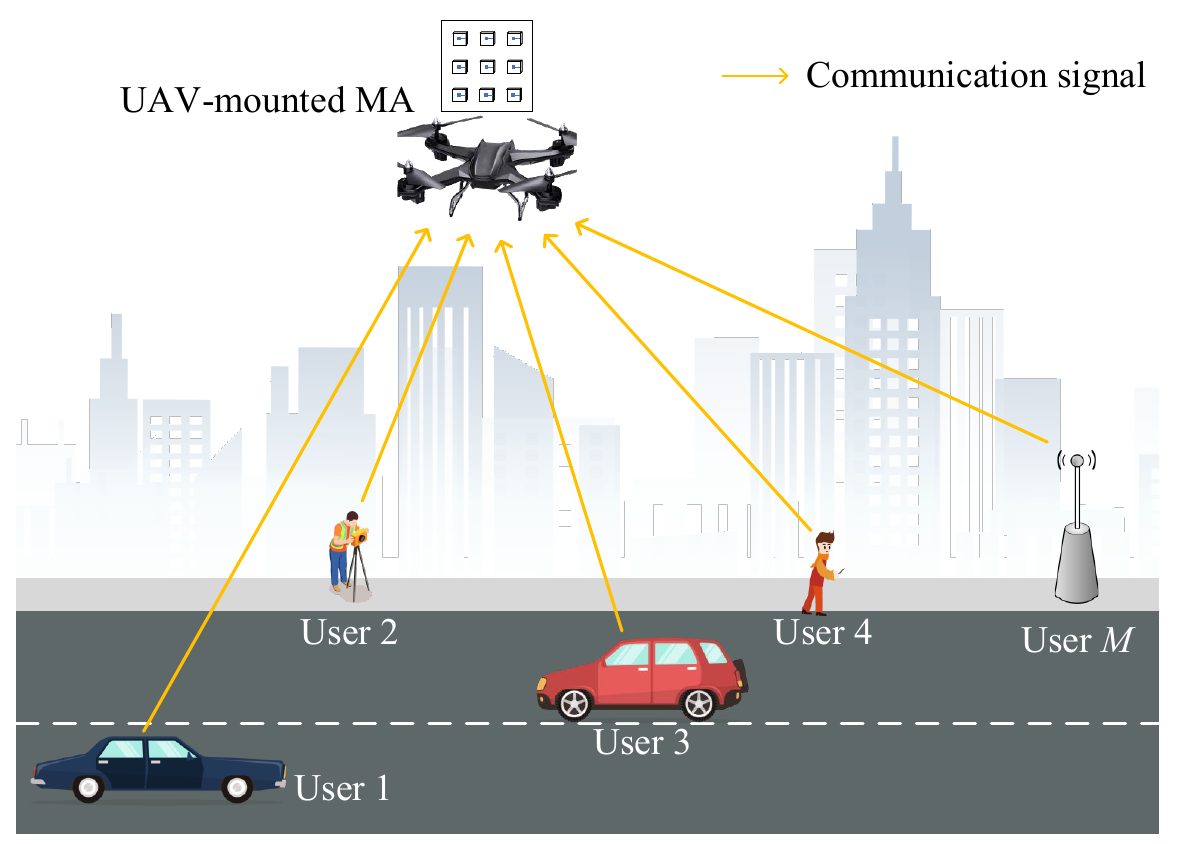}}
	\caption{The system model.}
	\label{fig. system model}
\end{figure}

As shown in Fig. \ref{fig. system model}, an MA-enabled uplink communication system is considered,
where an AAV is dispatched to fly over a designated region and collect data from $M$ ground users.
Each user is equipped with a single FPA, and the set of users is denoted as $\mathcal{M} \triangleq \{1,2, \ldots, M \}$.
The AAV is equipped with $K$ receive MAs, whose position can be flexibly adjusted within a given two-dimensional plane.\footnote{Noting that we assume that all antennas are placed vertically to the horizontal plane \cite{Xu2020Multiuser}.
Without loss of generality, we consider that the AAV's flight direction is independent of the vertical and horizontal angle-of-arrival (AoA) of the radiated communication signal beams, thereby simplifying the design of the MA position.}
All users share the same bandwidth on the uplink transmission to send data to the AAV, while also considering the mutual interference between users. 
The system operates within a designated mission period $T$, which is
divided into $N$ time slots, each with a duration of $\tau = \frac{T}{N}$, and the set of the time slots is denoted as $\mathcal{N} \triangleq \{1, \ldots, N\}$.

A three-dimensional Cartesian coordinate system is utilized to represent the position of the AAV and the users, where 
the location of the $m$-th user is denoted by ${\mathcal{P}}_m = \left[ \mathbf{s}_{m}, 0 \right]$, where $\mathbf{s}_{m} = \left[ x_{m}, y_{m} \right]$ represents the horizontal coordinates. 
The AAV flies at a fixed altitude $H$, which is sufficiently high to avoid any obstacles including terrains, towers, and buildings.
Additionally, given the short duration of each time slot, the AAV can be regarded as quasi-static within each time slot. 
Consequently, the position of the AAV at time slot $n$ is denoted as ${\mathbf{Q}}_n = [\mathbf{q}_{n}, H]$, where $\mathbf{q}_{n} = [x_{n}, y_{n}]$ is the horizontal coordinates.
The AAV departs from the starting position $\mathbf{q}_{\rm{I}}$ at the beginning of each mission period, and returns to the end position $\mathbf{q}_{\rm{F}}$ at the end of each mission period. 
During the entire flight, the AAV complies with the mobility constraints, with the maximum velocity $V_{\rm{max}}$ and maximum acceleration $a_{\rm{max}}$. 
Thus, the mobility constraints can be given by
\begin{equation}
    \mathbf{q}_{1}= \mathbf{q}_{\rm{I}}, \quad
    \mathbf{q}_{N}=\mathbf{q}_{\rm{F}},
\end{equation}
\begin{equation}
	\| \mathbf{v}_{n} \| =  \frac{\left \Vert \mathbf{q}_{n} - \mathbf{q}_{n-1} \right \Vert}{\tau}   \le V_{\rm{max}},\ \forall n\in \mathcal{N}\backslash\{1\},
\end{equation}
\begin{equation}
	\| \mathbf{a}_{n} \| =  \frac{\left \Vert \mathbf{v}_{n} - \mathbf{v}_{n-1} \right \Vert}{\tau}   \le a_{\rm{max}},\ \forall n\in \mathcal{N}\backslash\{1,2\}.
\end{equation}

We consider that the channel among the AAV and the users depends not only on the propagation environment but also on the position of the MA. 
Similarly to \cite{Qin2024Antenna, 10318061}, we adopt a far field-response based channel model, where the sizes of the MA moving region for the AAV is much smaller than the signal propagation distance.
Then, the angle-of-departure (AoD), the AoA, and the complex coefficient amplitude for each channel path between each user and the AAV remain constant regardless of the MA's position within the designated region, while only the phases of the multi-path channels vary.

Let $\mathbf{u}_{k,n} = \left[ x_{k,n}, y_{k,n} \right]^{\mathsf{}} \in \mathcal{R}_{s}$ denote the position of the $k$-th MA at the AAV in time slot $n$, where $\mathcal{R}_{s}$ is the square region with size $L \times L$ in planes. 
Then, the position vector of all $K$ MAs is denoted by $\mathbf{u}_{n} \triangleq \left[ \mathbf{u}_{1,n}^{\mathsf{}},\mathbf{u}_{2,n}^{\mathsf{}}, \cdots, \mathbf{u}_{K,n}^{\mathsf{}} \right]^{\mathsf{T}} \in \mathbb{R}^{K \times 2}$. 
To avoid potential electrical coupling between adjacent MAs, which could reduce antenna efficiency, a minimum inter-MA distance $d_{\min}$ is required between each pair of MAs at the AAV, i.e.
\begin{equation}
\| \mathbf{u}_{i,n} -\mathbf{u}_{j,n}\| \ge d_{\min}, \forall i \neq j, \forall i,j \in \mathcal{K}, \forall n \in \mathcal{N}.
\end{equation}

We assume that the transmission path between the $m$-th user and the AAV is $L_{m}^{t}$.
Then, we can use the $\theta_{m,i,n}$ and $\phi_{m,i,n}$ to represent the vertical AoA and the horizontal AoA of the $i$-th transmission path.
We assume that the AoA of all paths follows a uniform distribution in $\Delta$, which can be expressed as $\theta_{m,i,n} \in \left(\theta_{m,n}-\Delta / 2, \theta_{m,n}+\Delta / 2\right)$, and $\phi_{m,i,n} \in \left(\phi_{m,n}-\Delta / 2, \phi_{m,n}+\Delta / 2\right)$, where $\theta_{m,n}$ and $\phi_{m,n}$ are related with the locations of the $m$-th user and the AAV, which are given by
\begin{equation}
    \theta_{m,n}=\arcsin \frac{H}{d_{m,n}}, 
\end{equation}
\begin{equation}
        \varphi_{m,n}=\arccos \frac{ y_{n}-y_{m} }{\|\mathbf{q}_{n} - \mathbf{s}_{m} \|},
\end{equation}
where $d_{m,n} = \sqrt{\|\mathbf{q}_{n} - \mathbf{s}_{m} \|^{2}+H^2}$ represents the distance between the $m$-th user and the AAV.
Then, for the $m$-th user, the signal propagation phase difference of the $i$-th transmission path between the $k$-th MA position $\mathbf{u}_{k,n}$ and original position $\mathbf{u}_{0} = \left[0,0\right]^{\mathsf{T}}$ at the AAV can be given by
\begin{equation}
\begin{split}
       \rho_{m,k,i,n} = &\ x_{k,n}\sin \theta_{m,i,n}\cos\phi_{m,i,n} \\&+ y_{k,n}\sin\theta_{m,i,n}\sin\phi_{m,i,n}. 
\end{split}
\label{phase diffenent}
\end{equation}
Therefore, the corresponding receive field response vector (FRV) from the $m$-th user to the $k$-th MA at the AAV can be expressed as
\begin{equation}
 g_{m,k,n} =\left[{\mathrm{e}}^{{\mathrm{j}} \frac{2 \pi}{\lambda} \rho_{m,k,1,n}},  \ldots, {\mathrm{e}}^{{\mathrm{j}} \frac{2 \pi}{\lambda} \rho_{m,k,L_{m}^{t},n}}\right]^{\mathsf{T}} \in \mathbb{C}^{L_{m}^{t} \times 1},
\end{equation}
and the receive field response matrix can be represented as
\begin{equation}
\mathbf{G}_{m,n}=\left[g_{m,1,n}, g_{m,2,n}, \ldots, g_{m,K,n}\right] \in \mathbb{C}^{L_{m}^{t} \times K}.
\end{equation}
Since the user is equipped with single FPA, the transmit FRV for the $m$-th user is given by
\begin{equation}
   \mathbf{f}_{m}=[1,1, \ldots, 1]^{\mathsf{T}} \in \mathbb{R}^{L_{m}^{t} \times 1}.
\end{equation}

In addition, we define a path response matrix (PRM) $\mathbf{\Sigma}_{m,n} = \mathrm{diag} \left(\sigma_{m,1,n},\ldots,\sigma_{m,L_{m}^{t},n}  \right) \in \mathbb{C}^{L_{m}^{t} \times L_{m}^{t}}$, where $\sigma_{m,i,n}$ for $i = 1,2,\ldots,L_{m}^{t}$ represents the complex amplitude response for the $i$-th transmission path between the $m$-th user and the AAV.
The complex amplitude response $\sigma_{m,i,n}$ can be expressed as $\sigma_{m,i,n} = \sqrt{\frac{\alpha_{m,n}}{L_{m}^{t}}} \mathrm{g}_{m,i,n} $, where $\alpha_{m,n}$ denotes the large-scale fading channel gain and $\mathrm{g}_{m,i,n}$ denotes the small-scale fading coefficients \cite{10901248}.
The large-scale fading channel gain $\alpha_{m,n}$ can be expressed as $\alpha_{m,n} = h_0 d_{m,n}^{-2}$, where $h_0$ denotes the channel power gain at a reference distance of $\rm{1m}$ \cite{Xu2020Multiuser}.
Thus, the channel vector between the $m$-th user and the AAV can be expressed as
\begin{equation}
\begin{split}
\mathbf{h}_{m,n} & =  \mathbf{G}_{m,n}^{\mathsf{H}} \mathbf{\Sigma}_{m,n} \mathbf{f}_{m}  \\
& = \left[ \ \overline{h}_{m,1,n},\overline{h}_{m,2,n},\ldots,
\overline{h}_{m,K,n} \right]^{\mathsf{T}} \in \mathbb{C}^{K \times 1}. \label{channel_vector}
\end{split}
\end{equation}
where $\overline{h}_{m,k,n} = \sum_{i=1}^{L_{m}^{t}} \sigma_{m,i,n} e^{-j \frac{2 \pi}{\lambda} \rho_{m,k,i,n}} $.

Hence, the received signal from the $m$-th user is given by
\begin{equation} 
\begin{split}
&\mathbf{z}_{m,n}  = \mathbf{w}_{m,n}^{\mathsf{H}} \left ( \sum_{r\in \mathcal{M}} \sqrt{p_{r,n}}  
     \mathbf{h}_{r,n}s_{r,n} + \Tilde{\mathbf{n}}_{n} \right ) \\
& = \underbrace{ \mathbf{w}_{m,n}^{\mathsf{H}} \sqrt{p_{m,n}}  \mathbf{h}_{m,n}s_{m,n} }_{\rm{desired \ signal}}  + 
\underbrace{ \sum\limits_{ r = 1, r \neq m }^{ \mathcal{M} } \mathbf{w}_{m,n}^{\mathsf{H}} \sqrt{p_{r,n}} \mathbf{h}_{r,n}s_{r,n} }_{\rm{multiuser\ interference}} \\
& \quad + \mathbf{w}_{m,n}^{\mathsf{H}}\Tilde{\mathbf{n}}_{n},
\end{split} 
\end{equation}
where $\mathbf{w}_{m,n}$ denotes the receive beamforming vector of the AAV, $s_{r,n}$ denotes the information symbol for the $r$-th user which satisfies $\mathbb{E}[ \vert s_{r,n} \vert^{2} ] = 1$,
$p_{r,n}$ denotes the transmit power of the $r$-th user, $\Tilde{\mathbf{n}}_{n} \in \mathcal{C N}\left(\mathbf{0}, \sigma_{\mathrm{AAV}}^{2} \mathbf{I}_{K}\right)$ is the noise vector at the AAV. 
Thus, the received signal-to-interference-plus-noise ratio (SINR) of the AAV in time slot $n$ is given by
\begin{equation} 
\begin{split}
\gamma_{m,n} = 
 \cfrac{ p_{m,n} \left|  \mathbf{w}_{m,n}^{\mathsf{H}}
 \mathbf{h}_{m,n}  \right|^{2}  }
{ \sum\limits_{ r = 1, r \neq m }^{ \mathcal{M} } 
p_{r,n} \left|  \mathbf{w}_{m,n}^{\mathsf{H}} \mathbf{h}_{r,n}  \right|^{2} + \| \mathbf{w}_{m,n}^{\mathsf{H}} \|^{2}_{2}
\sigma_{\mathrm{AAV}}^{2}} .
\end{split}
\end{equation}
Then, the uplink transmission rate for the $m$-th user can be given by $R_{m,n} = \log_2 ( 1 + \gamma_{m,n} )$.

\subsection{Problem Formulation}
In this work, we focus on maximizing the sum-rate for all users under receive power constraints.
This is accomplished through the joint optimization of the AAV trajectory $\mathbf{Q} \triangleq \{\mathbf{q}_{n}, \forall n\}$, the receive beamforming $\mathbf{W} \triangleq \{\mathbf{w}_{m,n}, \forall m,\forall n \}$, the MA position $\mathbf{U} \triangleq \{\mathbf{u}_{k,n}, \forall k, \forall n\}$ and the user transmit power $\mathbf{P} \triangleq \{ p_{m,n}, \forall m ,\forall n\}$\footnote{
Given that the AAV's flight energy consumption significantly exceeds that of the MA movement and data collection, this work prioritizes enhancing the user data collection performance by optimizing the communication resource allocation and AAV trajectory design. Consequently, the energy consumption regarding the AAV aerodynamics will be addressed in our future works.
}. 
Therefore, the optimization problem can be formulated as
\begin{subequations} 
\begin{flalign}
 (\textbf{P1}):\ & \max_{\mathbf{Q},\mathbf{W},\mathbf{U},\mathbf{P} } \quad 
 \sum_{n=1}^{N} \sum_{m=1}^{M}  R_{m,n}  \label{p1o}
 \\
 {\rm{s.t.}}  \quad &\mathbf{q}_{1}= \mathbf{q}_{\rm{I}}, \quad
\mathbf{q}_{N}=\mathbf{q}_{\rm{F}}, \label{p1a} \\
 & \| \mathbf{v}_{n} \|  \le V_{\rm{max}}, \forall n\in \mathcal{N}\backslash\{1\}, \label{p1b} \\
 & \| \mathbf{a}_{n} \|   \le a_{\rm{max}}, \forall n\in \mathcal{N}\backslash\{1,2\},  \label{p1c} \\ 
 &  \Vert \mathbf{w}_{m,n} \Vert^2 \le 1 ,\ \forall m\in \mathcal{M}, \forall n\in \mathcal{N}, \label{p1d}  \\
 & \| \mathbf{u}_{i,n} -\mathbf{u}_{j,n}\| \ge d_{\min}, \forall i \neq j, \forall i,j \in \mathcal{K}, \forall n \in \mathcal{N}, \label{p1e} \\
 & \mathbf{u}_{k,n} \in \mathcal{R}_{s}, \forall k\in \mathcal{K}, \forall n\in \mathcal{N}, \label{p1f} \\
 & 0 \le  p_{m,n} \le P_{\rm{max}}, \forall m\in \mathcal{M}, \forall n\in \mathcal{N}, \label{p1g} 
\end{flalign}
\end{subequations}
\noindent where constraints (\ref{p1a})-(\ref{p1c}) represent the AAV kinematic constraints, (\ref{p1d}) represents the feasible region constraint of the receive beamforming, (\ref{p1e}) is the minimum distance constraint between adjacent MAs, (\ref{p1f}) represents the MA mobility constraint, (\ref{p1g}) denotes the transmit power constraint of the users.

\begin{remark}
    Due to the extremely complex form in which the MA position appear in Eqs. \eqref{phase diffenent} and \eqref{channel_vector} for the channel vectors between the users and the AAV, the uplink transmission rates exhibit a highly non-convex dependency on the MA position. Additionally, when the MA position are fixed, there is a strong coupling between the AAV trajectory and the beamforming variables. Furthermore, the optimization objective of Problem (\textbf{P1}) is non-concave, making the direct search for an optimal solution extremely difficult. Therefore, we propose an AO algorithm based on BCA, which seeks a suboptimal solution by alternately optimizing the beamforming, the AAV trajectory, the MA position, and the user transmit power. 
\end{remark}

\section{Proposed Alternating Optimization Algorithm}

\subsection{AAV Trajectory Optimization}
Given feasible receive beamforming $\mathbf{W}$, MA position $\mathbf{U}$ and user transmit power $\mathbf{P}$, the subproblem of optimizing the AAV trajectory $\mathbf{Q}$ is formulated as follows
\begin{subequations} 
\begin{flalign}
  (\textbf{P2}):\ & \max_{\mathbf{Q}} \  \sum_{n=1}^{N} \sum_{m=1}^{M}  R_{m,n} \label{p2a}  \\
  {\rm{s.t.}}  \quad &\eqref{p1a}-\eqref{p1c}.
\end{flalign}
\end{subequations}
Clearly, the objective function (\ref{p2a}) is non-convex with respect to $\mathbf{q}$. 
The variable $\mathbf{q}$ is included in both the received FRV and the PRM, which makes solving the problem (\textbf{P2}) optimally very challenging. 
To simplify the trajectory design, we use the AAV trajectory from the $l$-th iteration to approximate the receive FRV $g_{m,k,n}$ in the $(l+1)$-th iteration\cite{Deng2023Beamforming}.
As a result, the approximate receive FRV is given as
\begin{equation}
 \Tilde{g}_{m,k,n} =\left[e^{j \frac{2 \pi}{\lambda} \Tilde{\rho}_{m,k,1,n}},  \ldots, e^{j \frac{2 \pi}{\lambda} \Tilde{\rho}_{m,k,L_{m}^{t},n}}\right]^{\mathsf{T}},
\end{equation}
where 
\begin{equation} \small
   \Tilde{\rho}_{m,k,i,n} = x_{k,n}\sin \theta_{m,i,n}^{(l)}\cos\phi^{(l)}_{m,i,n} + y_{k,n}\sin\theta^{(l)}_{m,i,n}\sin\phi^{(l)}_{m,i,n}.
\end{equation}
Additionally, to ensure the accuracy of this approximation, we introduce a trust region constraint as\cite{9916163} 
\begin{equation}
    \| \mathbf{q}_{n}^{(l)} - \mathbf{q}_{n}^{(l-1)} \| \le \phi^{l}, \forall n \in \mathcal{N},
\end{equation}
where $\phi^{l}$ denotes the radius of the trust region in the $l$-th iteration.

Next, we focus only on the part of the PRM that is related to $\mathbf{q}$, i.e.
\begin{equation}
\begin{split}
\mathbf{\Sigma}_{m,n} & = \mathrm{diag} \left(\sigma_{m,1,n},\ldots,\sigma_{m,L_{m}^{t},n}  \right) \\
& = \sqrt{\frac{\alpha_{m,n}}{L_{m}^{t}}} 
\mathrm{diag} \left(\mathrm{g}_{m,1,n},\ldots,\mathrm{g}_{m,L_{m}^{t},n}   \right)\\
& = \frac{1}{d_{m,n}}
\mathrm{diag} \left(\sqrt{\frac{h_0}{L_{m}^{t}}}\mathrm{g}_{m,1,n},\ldots,\sqrt{\frac{h_0}{L_{m}^{t}}}\mathrm{g}_{m,L_{m}^{t},n}   \right).
\end{split}
\end{equation}
Thus, the channel vector from the $m$-th user to the AAV can be reformulated as
\begin{equation}
\begin{split}
\Tilde{\mathbf{h}}_{m,n} = \frac{1}{d_{m,n}}  \left[ \Tilde{h}_{m,1,n},\Tilde{h}_{m,2,n},\ldots,\Tilde{h}_{m,K,n} \right]^{\mathsf{T}} . 
\end{split}
\end{equation}
where $\Tilde{h}_{m,k,n} = \sqrt{\frac{h_0}{L_{m}^{t}}} \sum_{i=1}^{L_{m}^{t}} \mathrm{g}_{m,i,n} e^{-j \frac{2 \pi}{\lambda} \Tilde{\rho}_{m,k,1,n}} $. 
For simplicity, we let 
\begin{equation}
  \mathbf{h}_{m,n}^{\rm{\Xi}} =   \left[ \Tilde{h}_{m,1,n},\Tilde{h}_{m,2,n},\ldots,\Tilde{h}_{m,K,n} \right]^{\mathsf{T}} ,
\end{equation}
and $\Tilde{\mathbf{h}}_{m,n} = \frac{1}{d_{m,n}} \mathbf{h}_{m,n}^{\rm{\Xi}}$.
Then, the uplink transmission rate for the $m$-th user can be reformulated as 
\begin{equation} \small
\begin{split}
 &R_{m,n} \\
 &=  \log_2 \left( 1 + \cfrac{ d_{m,n}^{-2} p_{m,n} \left|  \mathbf{w}_{m,n}^{\mathsf{H}}
 \mathbf{h}_{m,n}^{\rm{\Xi}}  \right|^{2}  }
{ \sum\limits_{ r = 1, r \neq m }^{ \mathcal{M} } d_{r,n}^{-2}
p_{r,n} \left|  \mathbf{w}_{m,n}^{\mathsf{H}} \mathbf{h}_{r,n}^{\rm{\Xi}}  \right|^{2} + \| \mathbf{w}_{m,n}^{\mathsf{H}} \|^{2}_{2}
\sigma_{\mathrm{AAV}}^{2} } \right)   \\
&= \log_2 \left( \sum\limits_{ r = 1 }^{ \mathcal{M} }  d_{r,n}^{-2}
p_{r,n} \left|  \mathbf{w}_{m,n}^{\mathsf{H}} \mathbf{h}_{r,n}^{\rm{\Xi}} \right|^{2} + \| \mathbf{w}_{m,n}^{\mathsf{H}} \|^{2}_{2}
\sigma_{\mathrm{AAV}}^{2} \right) \\
& \quad - \log_2 \left( \sum\limits_{ r = 1, r \neq m }^{ \mathcal{M} } d_{r,n}^{-2}
p_{r,n} \left|  \mathbf{w}_{m,n}^{\mathsf{H}} \mathbf{h}_{r,n}^{\rm{\Xi}}  \right|^{2} + \| \mathbf{w}_{m,n}^{\mathsf{H}} \|^{2}_{2}
\sigma_{\mathrm{AAV}}^{2} \right) .
\end{split}
\end{equation}
Then, the first term of $R_{m,n}$ can be reformulated as
\begin{equation} \small
\begin{split}
& \quad \log_2 \left( \sum\limits_{ r = 1 }^{ \mathcal{M} }  d_{r,n}^{-2}  p_{r,n} \left|  \mathbf{w}_{m,n}^{\mathsf{H}} \mathbf{h}_{r,n}^{\rm{\Xi}}  \right|^{2} + \| \mathbf{w}_{m,n}^{\mathsf{H}} \|^{2}_{2}
\sigma_{\mathrm{AAV}}^{2} \right) \nonumber \\
& \ge \log_2 \left( \sum\limits_{ r = 1 }^{ \mathcal{M} }  (d_{r,n}^{(l)})^{-2}
p_{r,n} \left|  \mathbf{w}_{m,n}^{\mathsf{H}} \mathbf{h}_{r,n}^{\rm{\Xi}}  \right|^{2} + \| \mathbf{w}_{m,n}^{\mathsf{H}} \|^{2}_{2}
\sigma_{\mathrm{AAV}}^{2} \right) \nonumber \\
& + \sum\limits_{ r = 1 }^{ \mathcal{M} } E_{r,n} \left( (d_{r,n}^{(l)})^{2} - d_{r,n}^{2}  \right) 
\triangleq \hat{R}_{m,n}^{{ \rm first}} ,        
\end{split}
\end{equation}
where $d_{r,n}^{(l)} = \sqrt{\|\mathbf{q}^{(l)}_{n} - \mathbf{s}_{r} \|^{2}+H^2}$ represents the distance between the AAV and the user $r$ in the $l$-th iteration, and 
\begin{equation} \small
    \begin{split}
   E_{r,n} = \frac{ \log_{2}(e) (d_{r,n}^{(l)})^{-4} \left( \sum\limits_{ r = 1 }^{ \mathcal{M} } p_{r,n} \left|  \mathbf{w}_{m,n}^{\mathsf{H}} \mathbf{h}_{r,n}^{\rm{\Xi}}  \right|^{2} \right)   }{
\left( \sum\limits_{ r = 1 }^{ \mathcal{M} } (d_{r,n}^{(l)})^{-2}
p_{r,n} \left|  \mathbf{w}_{m,n}^{\mathsf{H}} \mathbf{h}_{r,n}^{\rm{\Xi}}  \right|^{2} \right)  + \| \mathbf{w}_{m,n}^{\mathsf{H}} \|^{2}_{2}
\sigma_{\mathrm{AAV}}^{2}
} .        
    \end{split}
\end{equation}

Since $d_{m,n}$ is convex with respect to $\mathbf{q}$, $\hat{R}_{m,n}^{{ \rm first}}$ is concave with respect to $\mathbf{q}$.
Moreover, the second term of $R_{m,n}$ is non-convex with respect to $\mathbf{q}$. To convert it to convex form, we introduce the slack variable $\eta_{m,n}$, which satisfies
\begin{equation}
\begin{split}
\frac{1}{{\rm e}^{\eta_{m,n}}} \le (d^{(l)}_{m,n})^{2} + 2(\mathbf{q}_{n}^{(l)} - & \mathbf{s}_{m})^{\rm{T}} (\mathbf{q}_{n}-\mathbf{q}^{(l)}_{n} ),\\
&\forall m \in \mathcal{M}, \forall n \in \mathcal{N}. \label{p2b}
\end{split}
\end{equation}
Then, the second term of $R_{m,n}$ can be reformulated as
\begin{equation} \small
\begin{split}
&\quad \log_2 \left( \sum\limits_{ r = 1, r \neq m }^{ \mathcal{M} } d_{r,n}^{-2}
p_{r,n} \left|  \mathbf{w}_{m,n}^{\mathsf{H}} \mathbf{h}_{r,n}^{\rm{\Xi}}  \right|^{2} + \| \mathbf{w}_{m,n}^{\mathsf{H}} \|^{2}_{2}
\sigma_{\mathrm{AAV}}^{2} \right)  \\
& \le   \log_2 \left( \sum\limits_{ r = 1, r \neq m }^{ \mathcal{M} }   {\rm e}^{\eta_{r,n}}
p_{r,n} \left|  \mathbf{w}_{m,n}^{\mathsf{H}} \mathbf{h}_{r,n}^{\rm{\Xi}}  \right|^{2} + \| \mathbf{w}_{m,n}^{\mathsf{H}} \|^{2}_{2}
\sigma_{\mathrm{AAV}}^{2} \right) \\
& \triangleq \check{R}_{m,n}^{{ \rm second}} .
\end{split}
\end{equation}

By replacing $R_{m,n}$ as its lower bound $\hat{R}_{m,n}^{{ \rm first}} - \check{R}_{m,n}^{{ \rm second}}$, the AAV trajectory optimization problem is approximated as the form below
\begin{subequations} 
\begin{flalign}
 (\textbf{P2-1}):\ & \max_{\mathbf{Q},\mathbf{\eta}} \quad
 \sum_{n=1}^{N} \sum_{m=1}^{M} ( \hat{R}_{m,n}^{{ \rm first}} - \check{R}_{m,n}^{{ \rm second}} ) \label{p2_1obj} \\
 {\rm{s.t.}}  \quad &\rm{(\ref{p1a})-(\ref{p1c}), (\ref{p2b})},
\end{flalign}
\end{subequations}
where $\mathbf{\eta} = \{ \eta_{m,n},\forall m ,\forall n \}$.
Problem (\textbf{P2-1}) is a convex optimization problem, which can
be efficiently solved by standard convex optimization solvers such as CVX toolbox. The detailed procedure to solve problem (\textbf{P2-1}) is summarized in Algorithm \ref{Algtra}.

\begin{algorithm} \footnotesize
	\caption{SCA-based Algorithm for Solving \textbf{P2-1}}
	\label{Algtra}
	\begin{algorithmic}[1]
		\REQUIRE {An initial feasible solution of trajectory ${\bf{Q}}^{l}$;}\\
		\textbf{Initialize:} the iteration number $l=0$;\\
		\textbf{Initialize:} the small precision threshold $\epsilon$;\\
            \textbf{Initialize:} the maximum number of iterations $l_{\max}$;\\
		
		\REPEAT
		\STATE Solve the problem \textbf{P2-1}, get the solution $\bold{Q}^{l} $;\\
		\STATE  Update the values of optimization variables;\\
		\STATE Compute the objective function $\Bar{R}^{l}$ via \eqref{p2_1obj};\\
		\STATE Set $l=l+1$; \\
		\UNTIL{$|\Bar{R}^{l}-\Bar{R}^{l-1}|$$<$$\epsilon$} or $l = l_{\max}$;\\
		\ENSURE {$\bold{Q}^{l},\Bar{R}^{l}$.}\\
	\end{algorithmic}
\end{algorithm}

\subsection{Receive Beamforming and User Transmit Power}
Given feasible AAV trajectory $\mathbf{Q}$ and the MA position $\mathbf{U}$, the subproblem of optimizing the receive beamforming $\mathbf{W}$ and the user transmit power $\mathbf{P}$ is formulated as follows
\begin{subequations} 
\begin{flalign}
 (\textbf{P3}):\ & \max_{\mathbf{W}, \mathbf{P}} \quad
\sum_{n=1}^{N} \sum_{m=1}^{M}  R_{m,n} \label{p3a}  \\
 {\rm{s.t.}}  \quad &\eqref{p1e}, \eqref{p1g}.
\end{flalign}
\end{subequations}

Problem (\textbf{P3}) poses significant challenges owing to the non-convex nature of the constraints delineated in (\ref{p3a}).  Furthermore, there exists a strong coupling between the user transmit power $\mathbf{P}$ and the beamforming $\mathbf{W}$, which complicates the problem further.
To make problem $(\textbf{P3})$ more tractable, we employ the weighted mean square error (WMMSE) method\cite{Shi2011An} to equivalently reformulate $R_{m,n}$ in the constraint (\ref{p3a}), as depicted in (\ref{p4a}). 
\begin{figure*}[!t]
\vspace*{-\baselineskip} 
\begin{equation} \small
\begin{aligned} 
&R_{m,n}  = \max\limits_{\omega_{m,n}\geq 0}  \log_{2}(\omega_{m,n}) - \omega_{m,n} \left( \sum\limits_{r=1}^{ \mathcal{M}} p_{r,n} \left|  \mathbf{w}_{m,n}^{\mathsf{H}} \mathbf{h}_{r,n}  \right|^{2} + \| \mathbf{w}_{m,n}^{\mathsf{H}} \|^{2}_{2}
\sigma_{\mathrm{AAV}}^{2}  \right)^{-1} \sqrt{p_{m,n}}  \mathbf{w}_{m,n}^{\mathsf{H}} \mathbf{h}_{m,n} + 1 \\
& =  \max\limits_{\omega_{m,n}\geq 0,\beta_{m,n}}  \log_{2}(\omega_{m,n}) - 
\omega_{m,n} \left(  1 - 2{\rm Re} \{ \beta_{m,n}^{*} \sqrt{p_{m,n}}  \mathbf{w}_{m,n}^{\mathsf{H}} \mathbf{h}_{m,n} \} + 
\left| \beta_{m,n} \right|^{2} \left(\sum\limits_{r=1}^{ \mathcal{M}} p_{r,n} \left|  \mathbf{w}_{m,n}^{\mathsf{H}} \mathbf{h}_{r,n}  \right|^{2} + \| \mathbf{w}_{m,n}^{\mathsf{H}} \|^{2}_{2}
\sigma_{\mathrm{AAV}}^{2} \right)  \right) +1 \\
&= \max\limits_{\omega_{m,n}\geq 0,\beta_{m,n}} \Tilde{R}_{m,n}.
 \label{p4a}
\end{aligned}
\end{equation}
\hrulefill
\end{figure*}

This transformation entails the introduction of auxiliary variables $\boldsymbol{\beta} =\{\beta_{m,n}, \forall m, \forall n\}$ and $\boldsymbol{\omega} =\{\omega_{m,n}, \forall m, \forall n\}$.
Therefore, the optimization problem is approximated in the following form
\begin{subequations} 
\begin{flalign}
 (\textbf{P3-1}):\ & \max_{\mathbf{W}, \mathbf{P},\boldsymbol{\beta}, \boldsymbol{\omega}} \quad
\sum_{n=1}^{N} \sum_{m=1}^{M}  \Tilde{R}_{m,n} \label{p3_1a}  \\
 {\rm{s.t.}}  \quad &\eqref{p1d}, \eqref{p1g},\\
  & \omega_{m,n}\geq 0, \forall m, \forall n.
\end{flalign}
\end{subequations}
Subsequently, we propose to utilize the BCA method to address the problem (\textbf{P3-1}).

\subsubsection{Updating Auxiliary Variables}
Based on the derivation of the WMMSE transformation, the auxiliary variables $\{\omega_{m,n}\}$ and $\{\beta_{m,n}\}$ can be updated using the following closed-form solutions, given the other variables:
\begin{equation}
\beta_{m,n}^{\rm opt} = \frac{\sqrt{p_{m,n}}  \mathbf{w}_{m,n}^{\mathsf{H}} \mathbf{h}_{m,n}}{ \sum\limits_{r=1}^{ \mathcal{M}} p_{r,n} \left|  \mathbf{w}_{m,n}^{\mathsf{H}} \mathbf{h}_{r,n}  \right|^{2} + \| \mathbf{w}_{m,n}^{\mathsf{H}} \|^{2}_{2}
\sigma_{\mathrm{AAV}}^{2} } , 
\label{p3beta}
\end{equation}
\begin{equation}
\omega_{m,n}^{\rm opt} = 1 + \frac{ p_{m,n} \left|  \mathbf{w}_{m,n}^{\mathsf{H}}
 \mathbf{h}_{m,n}  \right|^{2}  }
{ \sum\limits_{ r = 1, r \neq m }^{ \mathcal{M} } 
p_{r,n} \left|  \mathbf{w}_{m,n}^{\mathsf{H}} \mathbf{h}_{r,n}  \right|^{2} + \| \mathbf{w}_{m,n}^{\mathsf{H}} \|^{2}_{2}
\sigma_{\mathrm{AAV}}^{2}} . 
\label{p3omega}
\end{equation}

\subsubsection{Updating Receive Beamforming}

\begin{subequations} 
\begin{flalign}
 (\textbf{P3-2}):\ & \max_{\mathbf{W}} \quad \sum_{n=1}^{N} \sum_{m=1}^{M}  \Tilde{R}_{m,n} \\
 {\rm{s.t.}}  \quad &\eqref{p1d}.
\end{flalign}
\end{subequations}
Problem $(\textbf{P3-2})$ is a convex optimization problem, which can be efficiently solved by standard convex optimization solvers
such as CVX.
Moreover, by introducing dual variables associated with the constraints (\ref{p1d}), we can derive a Lagrangian dual function and thus a closed optimal solution can be derived. Concrete information is shown in Theorem \ref{Theorem1}.

\begin{theorem}\label{Theorem1}
For problem $(\textbf{P3-2})$, the optimal receive beamforming can be expressed as
\begin{equation}
    \begin{split}
&\mathbf{w}_{m,n}^{\rm opt} = \\
&\frac{ \omega_{m,n} \sqrt{p_{m,n}} \mathbf{h}_{m,n} \beta_{m,n}^{*}  }{ \omega_{m,n} \left| \beta_{m,n} \right|^{2} \left( 
\sum\limits_{r=1}^{ \mathcal{M}} p_{r,n} \mathbf{h}_{m,n}^{*} \mathbf{h}_{m,n}^{\mathsf{T}} + \sigma_{\mathrm{AAV}}^{2} \mathbf{I}  \right) + \lambda_{m,n} \mathbf{I}  },
    \end{split}
\end{equation}
where $\{\lambda_{m,n}\}$ are the dual variables associated with the corresponding constraints (\ref{p1d}).

\begin{proof}
Please refer to Appendix \ref{appendice1}.
\end{proof}
\end{theorem} 

\subsubsection{Updating User Transmit Power}
\begin{subequations} 
\begin{flalign}
 (\textbf{P3-3}):\ & \max_{\mathbf{P}} \quad \sum_{n=1}^{N} \sum_{m=1}^{M}  \Tilde{R}_{m,n} \\
 {\rm{s.t.}}  \quad &\eqref{p1g}.
\end{flalign}
\end{subequations}
By treating $p_{m,n}^{\rm{new}} = \sqrt{p_{m,n}}$ as an optimization variable, we obtain a new optimization problem
\begin{subequations} 
\begin{flalign}
 (\textbf{P3-4}):\ & \max_{\mathbf{P^{\rm{new}}}} \quad \sum_{n=1}^{N} \sum_{m=1}^{M}  \Tilde{R}^{\rm{new}}_{m,n} \\
 {\rm{s.t.}}  \quad &0 \le  p_{m,n}^{\rm{new}} \le \sqrt{P_{\rm{max}}}, \forall m\in \mathcal{M}, \forall n\in \mathcal{N},\label{p3-3b}
\end{flalign}
\end{subequations}
where $\mathbf{P^{\rm{new}}} \triangleq \{ p_{m,n}^{\rm{new}}, \forall m ,\forall n\}$, and $\Tilde{R}^{\rm{new}}_{m,n}$ is given in Equation \eqref{p3-3a}.

\begin{figure*}[!t]
\vspace*{-\baselineskip} 
\begin{equation} \small
\begin{aligned} 
\Tilde{R}^{\rm{new}}_{m,n}  =  \log_{2}(\omega_{m,n}) - 
\omega_{m,n} \left(  1 - 2{\rm Re} \{ \beta_{m,n}^{*} p_{m,n}^{\rm{new}}  \mathbf{w}_{m,n}^{\mathsf{H}} \mathbf{h}_{m,n} \} + 
\left| \beta_{m,n} \right|^{2} \left(\sum\limits_{r=1}^{ \mathcal{M}} (p_{r,n}^{\rm{new}})^2 \left|  \mathbf{w}_{m,n}^{\mathsf{H}} \mathbf{h}_{r,n}  \right|^{2} + \| \mathbf{w}_{m,n}^{\mathsf{H}} \|^{2}_{2}
\sigma_{\mathrm{AAV}}^{2} \right)  \right) +1 
 \label{p3-3a}
\end{aligned}
\end{equation}
\hrulefill
\end{figure*}

Problem $(\textbf{P3-4})$ is a convex optimization problem, which can be efficiently solved by standard convex optimization solvers
such as CVX.
Moreover, by introducing dual variables associated with the constraints (\ref{p3-3b}), we can derive a Lagrangian dual function and thus an optimal solution can be derived. Concrete information is shown in Theorem \ref{Theorem2}.

\begin{theorem}\label{Theorem2}
For problem ${(\textbf{P3-4})}$, the optimal user transmit power can be expressed as
\begin{equation}
    \begin{split}
(p_{m,n}^{\rm{new}})^{\rm opt} = \frac{ -\mu_{m,n} + 2 \omega_{m,n} {\rm Re} \{ \beta_{m,n}^{*}   \mathbf{w}_{m,n}^{\mathsf{H}} \mathbf{h}_{m,n} \}  }{ 2 \sum_{r=1}^{M} \omega_{r,n} \left| \beta_{r,n} \right|^{2}  \left|  \mathbf{w}_{r,n}^{\mathsf{H}} \mathbf{h}_{m,n}  \right|^{2}  },
    \end{split}
\end{equation}
where $\{\mu_{m,n}\}$ are the dual variables associated with the corresponding constraints \eqref{p3-3b}.

\begin{proof}
Please refer to Appendix \ref{appendice2}.
\end{proof}
\end{theorem} 

The detailed BCA procedure to solve problem (\textbf{P3-1}) is summarized in Algorithm \ref{AlgBCA}.

\begin{algorithm} \footnotesize
	\caption{BCA-based Algorithm for Solving \textbf{P3-1}}
	\label{AlgBCA}
	\begin{algorithmic}[1]
		\REQUIRE {An initial feasible solution ${\bf{W}}^{0}$ and ${\bf{P}}^{0}$;}\\
		\textbf{Initialize:} the iteration number $j=0$;\\
		\textbf{Initialize:} the small precision threshold $\epsilon$;\\
            \textbf{Initialize:} the maximum number of iterations $j_{\max}$;\\
		\REPEAT
		\STATE Update the auxiliary variables and obtain the solution via \eqref{p3beta} and \eqref{p3omega};\\
		\STATE  Update the receive beamforming by solving \textbf{P3-2};\\
        \STATE  Update the user transmit power by solving \textbf{P3-4};\\
		\STATE Compute the objective function $\Bar{R}^{j}$ via \eqref{p3_1a};\\
		\STATE Set $j=j+1$; \\
		\UNTIL{$|\Bar{R}^{j}-\Bar{R}^{j-1}|$$<$$\epsilon$} or $j = j_{\max}$;\\
		\ENSURE {$\bold{W}^{j},\bold{P}^{j},\Bar{R}^{j}$.}\\
	\end{algorithmic}
\end{algorithm}

\subsection{Antenna Position of MAs}
Given the AAV trajectory $\mathbf{Q}$, the receive beamforming $\mathbf{W}$, and the user transmit power $\mathbf{P}$,
the problem can be simplified as the subproblem of optimizing the MA position $\mathbf{U}$, which is formulated as
\begin{subequations} 
\begin{flalign}
 (\textbf{P4-1}):\ & \max_{\mathbf{U}} \quad \sum_{n=1}^{N} \sum_{m=1}^{M}  {R}_{m,n} \\
 {\rm{s.t.}}  \quad &\eqref{p1e}-\eqref{p1f}.
\end{flalign}
\end{subequations}
The solutions of $\mathbf{U}$ are independent across different time slots.
To reduce the computational complexity, we decompose problem \textbf{P4-1} into $N$ subproblems corresponding to different time slots.
Hence, each subproblem can be optimized independently.
Specifically, the $n$-th subproblem in time slot $n$ can be formulated as
\begin{subequations} 
\begin{flalign}
 (\textbf{P4-1.}n):\ & \max_{\mathbf{U}_n} \quad  \sum_{m=1}^{M}  {R}_{m,n} \\
 {\rm{s.t.}}  \quad & \| \mathbf{u}_{i,n} -\mathbf{u}_{j,n}\| \ge d_{\min}, \forall i \neq j, \forall i,j \in \mathcal{K}, \label{p41na} \\
 & \mathbf{u}_{k,n} \in \mathcal{R}_{s}, \forall k\in \mathcal{K}, \label{p41nb}
\end{flalign}
\end{subequations}
where $\mathbf{U}_n$ denotes the solution of the antenna position in the $n$-th time slot.
Owing to the strongly non-convex nature of problem \textbf{P4-1} and high dimensionality of the solution space of the antenna position, the traditional optimization methods are hard to solve the problem \textbf{P4-1}, and incur impractical computational burdens under exhaustive search strategies.
Consequently, particle swarm optimization (PSO) is implemented to address this problem \cite{10741192, 10818453}.

Firstly, we initialize $S$ particles to represent the feasible antenna positions of the MAs.
Specifically, the particle set is denoted as $\boldsymbol{\mathcal{P}}^{(0)}_{n} = \{ \boldsymbol{\mathcal{P}}^{(0)}_{1,n}, \boldsymbol{\mathcal{P}}^{(0)}_{2,n}, \ldots, \boldsymbol{\mathcal{P}}^{(0)}_{S,n} \}$, where the $s$-th particle is given by
\begin{equation}
    \boldsymbol{\mathcal{P}}^{(0)}_{s,n} = \{ \bold{u}_{1,n}^{(0)}, \bold{u}_{2,n}^{(0)}, \ldots, \bold{u}_{K,n}^{(0)} \},
\end{equation}
where $\bold{u}_{k,n}^{(0)} = [x_{k,n}^{(0)}, y_{k,n}^{(0)}] \in \mathcal{R}_s $ denotes the feasible position of the $k$-th MA.
Meanwhile, the velocities of the particles are given by $\boldsymbol{\mathcal{V}}^{(0)}_n = \{ \boldsymbol{\mathcal{V}}^{(0)}_{1,n}, \boldsymbol{\mathcal{V}}^{(0)}_{2,n}, \ldots, \boldsymbol{\mathcal{V}}^{(0)}_{S,n} \}$.

Let $\Tilde{\boldsymbol{\mathcal{P}}}_{s,n}^{\mathrm{L}}$ denote the best position of the $s$-th particle, and $\Tilde{\boldsymbol{\mathcal{P}}}_{n}^{\mathrm{G}}$ denote the best position of the particle swarm.
In each iteration, the update of the velocity of the $s$-th particle can be given by
\begin{equation}
\begin{split}
    \boldsymbol{\mathcal{V}}_{s,n}^{(t+1)} \leftarrow
    \chi \boldsymbol{\mathcal{V}}_{s,n}^{(t)} &+
    \mathsf{L}_1 \mathsf{R}_1 \left( \Tilde{\boldsymbol{\mathcal{P}}}_{s,n}^{\mathrm{L}} - \boldsymbol{\mathcal{P}}_{s,n}^{(t)} \right)\\&+
    \mathsf{L}_2 \mathsf{R}_2 \left( \Tilde{\boldsymbol{\mathcal{P}}}_{n}^{\mathrm{G}} - \boldsymbol{\mathcal{P}}_{s,n}^{(t)} \right),
\end{split}
\label{pvel}
\end{equation}
where $t$ denotes the iteration index of the PSO.
The individual learning factor $\mathsf{L}_1$ and the global learning factor $\mathsf{L}_2$ determine the step size the particles take toward their individual best position and the swarm's best position, respectively.
To increase search randomness and avoid local optima, random variables $\mathsf{R}_1$ and $\mathsf{R}_2$ are sampled uniformly from $[0, 1]$, i.e., $\mathsf{R}_1, \mathsf{R}_2 \sim \mathcal{U} [0,1]$.
Momentum in particle motion is sustained by the inertia weight $\chi$.
Particularly, to balance the trade-off between convergence speed and precision, the inertia weight $\chi$ is dynamically decreased during iterations according to
\begin{equation}
    \chi = \chi_{\max} - \frac{
    \left ( \chi_{\max} - \chi_{\min} 
    \right )t
    }{
    t_{\max}
    },
    \label{inertia_weight}
\end{equation}
where $\chi_{\max}$ and $\chi_{\min}$ denote the maximum and minimum value of the inertia weight, respectively. $t_{\max}$ is the maximum iteration number of the PSO.
Meanwhile, the position of the $s$-th particle is update by the velocity, and also constrained by the array region, which can be given by
\begin{equation}
    \boldsymbol{\mathcal{P}}_{s,n}^{(t+1)} \leftarrow \min \left( \max \left( \left(
    \boldsymbol{\mathcal{P}}_{s,n}^{(t)} + \boldsymbol{\mathcal{V}}_{s,n}^{(t+1)}
    \right), 0 \right), L \right).
    \label{ppos}
\end{equation}

We evaluate the fitness function as $\Bar{\mathcal{R}} \left( \boldsymbol{\mathcal{P}}^{(t)}_{n} \right) = \sum_{n=1}^N \sum_{m=1}^M R_{m,n} (\mathbf{Q}, \mathbf{W}, \mathbf{P})$, for given the trajectory $\mathbf{Q}$, the receive beamforming $\mathbf{W}$, and the user transmit power $\mathbf{P}$.
Similarly, we introduce the adaptive penalty factor to the fitness function as
\begin{equation}
    \digamma \left( \boldsymbol{\mathcal{P}}^{(t)}_{n} \right) =
    \Bar{\mathcal{R}} \left( \boldsymbol{\mathcal{P}}^{(t)}_{n} \right) - \psi \left\vert \Bar{\mathcal{U}} \left( \boldsymbol{\mathcal{P}}^{(t)}_{n} \right) \right\vert,
    \label{PSOfitness}
\end{equation}
where $\psi$ denotes a large positive parameter serves as a penalty factor.
$\Bar{\mathcal{U}} \left( \boldsymbol{\mathcal{P}}^{(t)}_{n} \right)$ denotes the set of the MA position pairs violating the minimum distance constraint in \eqref{p1e}, which can be expressed as
\begin{equation}
\begin{split}
\Bar{\mathcal{U}} \left( \boldsymbol{\mathcal{P}}^{(t)}_{n} \right) = 
    \{
        (\bold{u}_{i,n},\bold{u}_{j,n})\vert
        \vert \bold{u}_{i,n}& - \bold{u}_{j,n} \vert_2
        < d_{\min},\\ &\ 1\leq i < j \leq K
        \}.
\end{split}
\end{equation}
The penalty factor is introduced to force the particle swarm to exclusively explore the valid positions to ensure the minimum MA distance constraint \eqref{p41na}, as infeasible solutions automatically receive fitness functions \eqref{PSOfitness} below zero.
The detailed PSO-based algorithm for addressing problem \textbf{P4-1} is summarized in Algorithm \ref{algorithmPSO}. 

\begin{algorithm}[t] \footnotesize
	\caption{The PSO-based Algorithm for Solving \textbf{P4-1}.}
		\begin{algorithmic}[1]
			\REQUIRE {An initial feasible solution ${\bf{U}}^{}$;}
            \FOR{$n$ $=1$ to $N$}
			\STATE
			\textbf{Initialize:} $S$ particles with initial positions $\boldsymbol{\mathcal{P}}_n^{(0)}$ and velocities $\boldsymbol{\mathcal{V}}_n^{(0)}$, the local best position of each particle, and the global best position of the swarm;
            \STATE
            Evaluate the initial fitness value;
            \FOR{$t$ $=1$ to $t_{\max}$}
                \STATE
                Update the inertia weight via \eqref{inertia_weight};
            \FOR{$s$ $=1$ to $S$}
                \STATE
                Update the velocity of the $s$-th particle via \eqref{pvel};
                \STATE
                Update the position of the $s$-th particle via \eqref{ppos};
                \STATE
                Calculate the fitness value via \eqref{PSOfitness};
                \IF{$\digamma \left( \boldsymbol{\mathcal{P}}^{(t)}_{s,n} \right) > \digamma \left( \Tilde{\boldsymbol{\mathcal{P}}}^{\mathrm{L}}_{s,n} \right)$}
                \STATE
                Update $\digamma \left( \Tilde{\boldsymbol{\mathcal{P}}}^{\mathrm{L}}_{s,n} \right) \leftarrow \digamma \left( \boldsymbol{\mathcal{P}}^{(t)}_{s,n} \right)$;
                \ENDIF
                \IF{$\digamma \left( \boldsymbol{\mathcal{P}}^{(t)}_{s,n} \right) > \digamma \left( \Tilde{\boldsymbol{\mathcal{P}}}^{\mathrm{G}}_{n} \right)$}
                \STATE
                Update $\digamma \left( \Tilde{\boldsymbol{\mathcal{P}}}^{\mathrm{G}}_{n} \right) \leftarrow \digamma \left( \boldsymbol{\mathcal{P}}^{(t)}_{s,n} \right)$;
                \ENDIF
            \ENDFOR
            \ENDFOR
            \ENDFOR
            \ENSURE
			{${\bf{U}}^{} (\boldsymbol{\mathcal{P}}^{(T)})$ given the position of the swarm $\boldsymbol{\mathcal{P}}^{(T)}_n,\ \forall n \in \mathcal{N}$}.
		\end{algorithmic}
        \label{algorithmPSO}
\end{algorithm}

\subsection{Algorithm Analysis}
We utilize an AO structure to solve the problem \textbf{P1}, and the overall procedure is summarized in Algorithm \ref{algoverall}.
The following proposition analyze the convergence of Algorithm \ref{algoverall}.
\begin{algorithm} \footnotesize
	\caption{AO Algorithm for Solving \textbf{P1}}
	\label{algoverall}
	\begin{algorithmic}[1]
		\REQUIRE {An initial feasible solution ${\bf{Q}}^{i}$, ${\bf{W}}^{i}$, ${\bf{W}}^{i}$, and ${\bf{P}}^{i}$;}\\
		\textbf{Initialize:} the iteration number $i=0$;\\
		\textbf{Initialize:} the small precision threshold $\epsilon$;\\
            \textbf{Initialize:} the maximum number of iterations $i_{\max}$;\\
		\REPEAT
		\STATE Solving problem \textbf{P2-1} via Algorithm \ref{Algtra}, get the solution ${\bf{Q}}^{i}$;\\
		\STATE Solving problem \textbf{P3-1} via Algorithm \ref{AlgBCA}, get the solution ${\bf{W}}^{i}$ and ${\bf{P}}^{i}$;\\
            \STATE Solving problem \textbf{P4-1} via Algorithm \ref{algorithmPSO}, get the solution ${\bf{U}}^{i}$;\\
		\STATE Compute the objective function $\Bar{R}^{i}$ via \eqref{p1o};\\
		\STATE Set $i=i+1$; \\
		\UNTIL{$|\Bar{R}^{i}-\Bar{R}^{i-1}|$$<$$\epsilon$} or $i = i_{\max}$;\\
		\ENSURE {The optimized solution $\bold{Q}^{i}$, $\bold{W}^{i}$, $\bold{P}^{i}$, and $\bold{U}^{i}$.}\\
	\end{algorithmic}
\end{algorithm}

\begin{proposition}
\label{pconvergence}
    The objective function of problem \textbf{P1} keeps increasing as the number of iterations increases. Therefore, Algorithm \ref{algoverall} is guaranteed to converge.
\begin{proof}
Please refer to Appendix \ref{propostionconv}.
\end{proof}
\end{proposition}

In Algorithm \ref{algoverall}, we employ the interior-point method to solve the three subproblems: AAV trajectory optimization, user transmit power control, and receive beamforming. The computational complexities of these subproblems are denoted by $\mathcal{O} ((2N)^{3.5} \log(\epsilon^{-1}))$, $\mathcal{O} ((MN)^{3.5} \log(\epsilon^{-1}))$, and $\mathcal{O} ((MNK^{2})^{3.5} \log(\epsilon^{-1})) $, respectively, where $\epsilon$ represents the stopping tolerance~\cite{Deng2023Beamforming, 5447068}.
Furthermore, the PSO algorithm is employed to solve for the MA positions, with a computational complexity denoted by $\mathcal{O} ( N t_{\max} S ( 2K + \log(S) ) )$~\cite{10741192}. 
Therefore, the total computational complexity of Algorithm \ref{algoverall} is $\mathcal{O} ( I_{\rm{1}} ( 
\mathcal{O} ((2N)^{3.5} \log(\epsilon^{-1})) +
I_{\rm{2}} (\mathcal{O} ((MN)^{3.5} \log(\epsilon^{-1})) + 
\mathcal{O} ((MN)^{3.5} \log(\epsilon^{-1})) ) + 
N t_{\max} S ( 2K + \log(S) ) 
) )$, where $I_{\rm{1}}$ denotes the resultant iteration number of Algorithm \ref{algoverall}, and $I_{\rm{2}}$ denotes the resultant iteration number of Algorithm \ref{AlgBCA}.

\section{Numerical Results}
In this section, we provide the numerical results of the proposed scheme with extensive simulation experiments.
We consider a square area of $800 \times 800 \ \rm{m}^{2}$ for the data collection mission.
The flying height of the AAV is $50\ \rm{m}$.
To evaluate the performance gain, we compare the proposed schemes with the following three benchmark schemes
\begin{itemize}
    \item AO-MM Scheme: We utilize the minorization
maximization (MM) method to optimize the antenna position, and the detailed procedures are presented in Appendix \ref{AOMM}.
    \item Fixed Trajectory Scheme: The trajectory of the AAV is fixed, while the receive beamforming and the antenna position are jointly optimized.
    \item FPA Scheme: The AAV is equipped with an FPA plane, while the trajectory and the receive beamforming are jointly optimized.
\end{itemize}
Meanwhile, unless otherwise stated, the default simulation parameters are summarized in Table \ref{tnoa}.

\begin{table}[!htbp]
\centering
\caption{Simulation Parameters.}
\label{tnoa}
\begin{tabular}{p{6.2cm}c}
\toprule
\centering
Parameters & Value\\
\midrule  %
\centering Mission period, $T$ &  40 s\\
\centering Number of time slot, $N$ & 20\\
\centering Duration of time slot, $\tau$ & 2 s\\
\centering Number of user, $M$ & 4\\
\centering Number of antenna, $K$ & 4\\
\centering Number of channel path for each user, $L_{m}^{t}$ & 4\\
\centering Carrier wavelength, $\lambda$ & 0.1 $\mathrm{m}$\\
\centering Size of antenna mobility region , $\mathcal{R}_{s}$ & $4\lambda \times 4\lambda$ \\
\centering Minimum distance between antennas, $d_{\rm{min}}$ & $0.5\lambda$ \\
\centering Rician factor, $\kappa$ & 15 \\
\centering Maximum speed, $V_{\rm{max}}$& $30\ {\mathrm{m/s}}$ \\
\centering Maximum acceleration, $a_{\rm{max}}$ & $10\ {\mathrm{m/s}}^{2}$ \\
\centering Maximum transmit power, $P_{\rm{max}}$ & 1 w\\
\centering Channel power gain at the reference distance $1 \mathrm{m}$, $h_{0}$& -60 dB\\
\centering Noise power at the AAV, $\sigma_{\rm{AAV}}$& -110 dBm\\
\centering Number of particles in PSO algorithm, $S$ & 100\\
\centering Maximum iterations in PSO algorithm, $t_{\max}$ & 100\\
\centering Individual learning factor in PSO algorithm, $\mathsf{L}_1$ & 1.4\\
\centering Global learning factor in PSO algorithm, $\mathsf{L}_2$ & 1.4\\
\centering Minimum inertial weight in PSO algorithm, $\chi_{\min}$ & 0.4\\
\centering Maximum inertial weight in PSO algorithm, $\chi_{\max}$ & 0.9\\
\centering Penalty factor in PSO algorithm, $\psi$ & 20\\
\bottomrule
\end{tabular}
\end{table}

\begin{figure}[!htbp]
	\centering
	\includegraphics[width=0.88\linewidth]{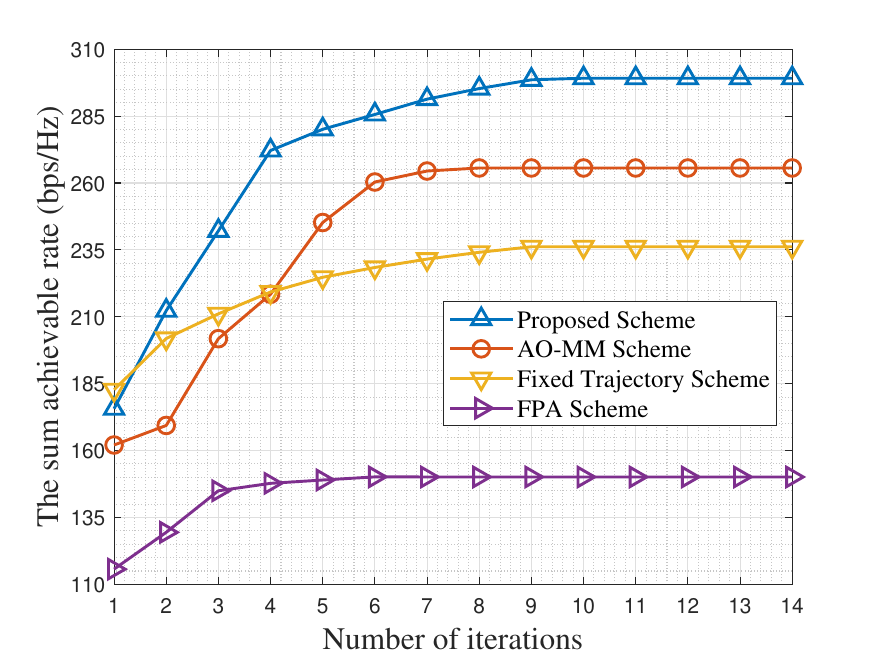}
	\caption{The convergence performance.}
	\label{convergence}
\end{figure}
Fig. \ref{convergence} demonstrates the convergence of the sum achievable rate across different schemes. The proposed scheme achieves over $300$ bps/Hz sum achievable rate within $9$ iterations, exhibiting the fastest convergence rate among all schemes.
Notably, the fixed trajectory scheme exhibits slower convergence and lower sum achievable rate compared to the proposed scheme and the AO-MM scheme, highlighting the vital role of trajectory optimization in overall system performance. Furthermore, the FPA scheme achieves the lowest sum achievable rate at convergence due to its inability to leverage the precise beamforming gains empowered by the MAs.

\begin{figure}[!htbp]
	\centering
	\includegraphics[width=0.88\linewidth]{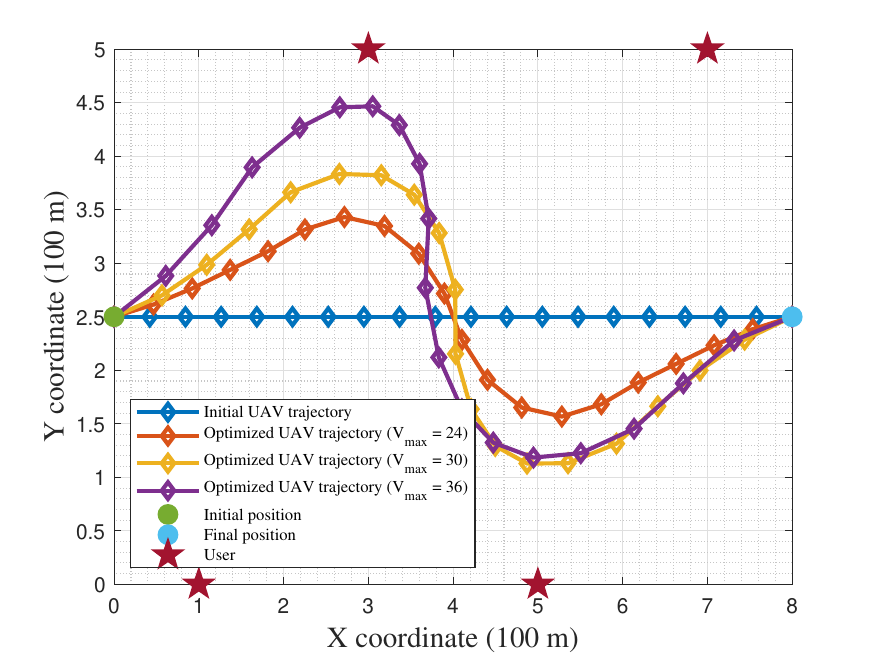}
	\caption{The optimized trajectory of the AAV.}
	\label{AAV_trajectory}
\end{figure}
Fig. \ref{AAV_trajectory} presents the optimized trajectories of the AAV under different maximum velocity of the AAV.
We can observe that a higher maximum AAV velocity enables the targeted flight trajectory toward the users with better channel conditions, where the MAs leverage precise beamforming to enhance power allocation while suppressing inter-user interference, which is capable of collectively boosting sum achievable rates of users.
Conversely, reduced maximum velocity constraints of the AAV prompt balanced trajectory designs ensuring equitable proximity to all users, maintaining optimized spectral efficiency under the AAV mobility limitations.

\begin{figure}[!htbp]
	\centering
	\includegraphics[width=0.88\linewidth]{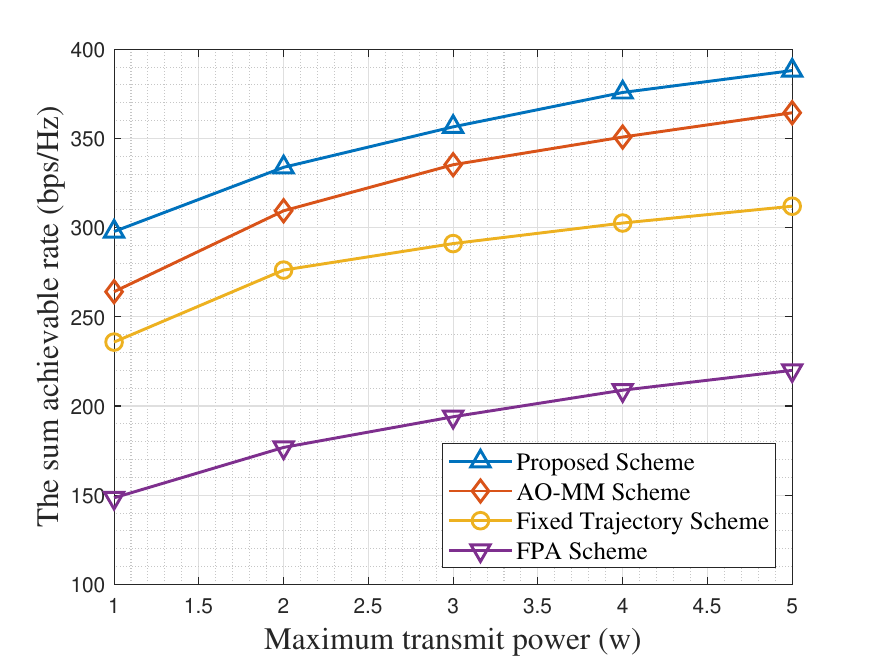}
	\caption{The sum achievable rate versus maximum
 transmit power.}
	\label{transmit_power}
\end{figure}
Fig. \ref{transmit_power} demonstrates that the sum achievable rate for all schemes increases with the maximum transmit power. This is attributed to the enhanced transmit power overcoming the path loss and improving the system SINR.
Furthermore, while increasing the maximum transmit power generally improves the sum achievable rate, the marginal gains diminish as the transmit power grows.
This saturation effect occurs because higher transmit power, although boosting the desired user's signal power, proportionally amplify the received interference from other co-channel users.
Additionally, the proposed scheme consistently outperforms other schemes. It achieves superior user-oriented beamforming through more effective antenna position optimization, leading to better interference management compared to the AO-MM scheme.
In contrast, both the fixed trajectory scheme and the FPA scheme suffer from inefficient beamforming gain, due to their inability to dynamically steer beams optimally towards users, resulting in substantially lower sum achievable rates across the transmit power range.

\begin{figure}[!htbp]
	\centering
	\includegraphics[width=0.88\linewidth]{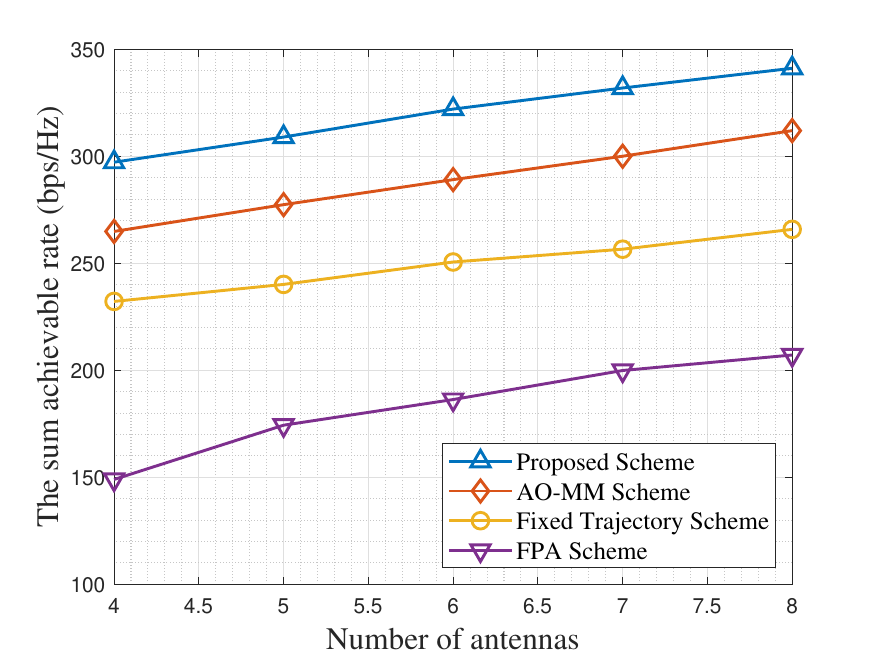}
	\caption{The sum achievable rate versus number of antennas.}
	\label{number_antenna}
\end{figure}
Fig. \ref{number_antenna} illustrates the sum achievable rate versus the number of receive antennas at the AAV.
We can observe that the increasing number of the receive antennas improves the sum achievable rate for all schemes, primarily due to the enhanced receive beamforming gains and spatial resolution.
The proposed scheme demonstrates significant advantage, leveraging its optimized antenna positioning to achieve more precise user-specific beam alignment.
This enables superior interference suppression compared to the AO-MM scheme, especially in multi-user scenarios.
Conversely, both the fixed trajectory scheme and the FPA scheme exhibit lower sum achievable rate performance. The reason is that, their static configurations cannot dynamically concentrate energy toward specific users, resulting in significantly lower sum achievable rate as the number of receive antennas increases.

\begin{figure}[!htbp]
	\centering
	\includegraphics[width=0.88\linewidth]{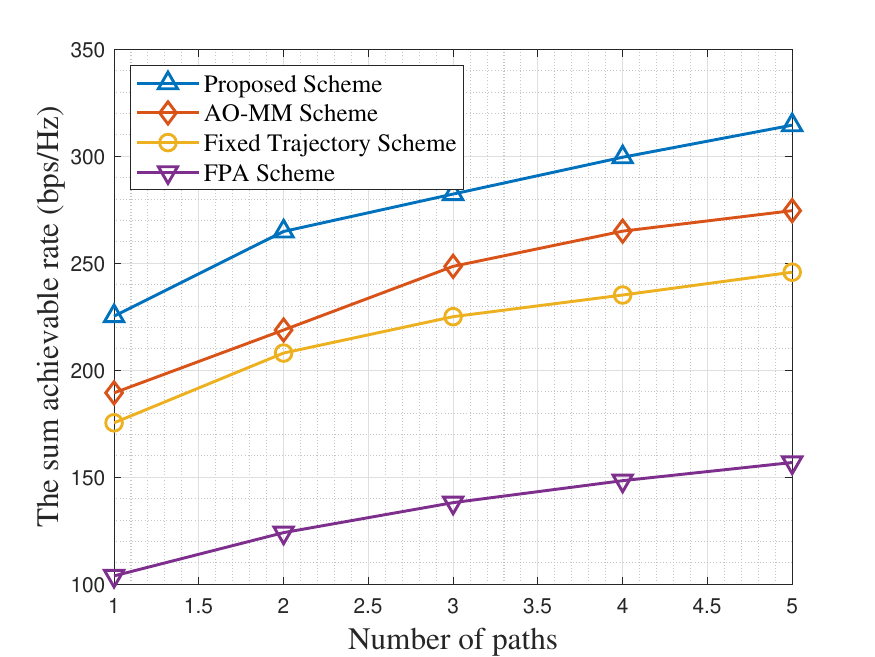}
	\caption{The sum achievable rate versus number of path.}
	\label{number_path}
\end{figure}
Fig. \ref{number_path} demonstrates that the
increasing multi-path components enhance the sum achievable rate across all schemes.
This can be attributed to the spatial diversity gain inherent in mutli-antenna systems, i.e., the richer scattering environments create additional spatial degrees of freedom, enabling more effective multi-path energy exploitation, and improving the sum achievable rate.
The proposed scheme achieves significantly steeper growth by dynamically optimizing antenna positions to coherently combine dominant propagation paths. This adaptive spatial matching maximizes the signal power at target users while suppressing uncorrelated noise.
The AO-MM scheme exhibits lower sum achievable rate gains due to the suboptimal antenna positions. 
The fixed trajectory scheme and the FPA scheme show limited improvements in the sum achievable rate. The reason is that their static configurations cannot exploit phase coherence across scattering paths, resulting in lower sum achievable rate compared to the proposed scheme across all paths. This highlights the critical role of dynamic antenna position and AAV maneuver control in harvesting multi-path benefits.

\begin{figure}[!htbp]
	\centering
	\includegraphics[width=0.88\linewidth]{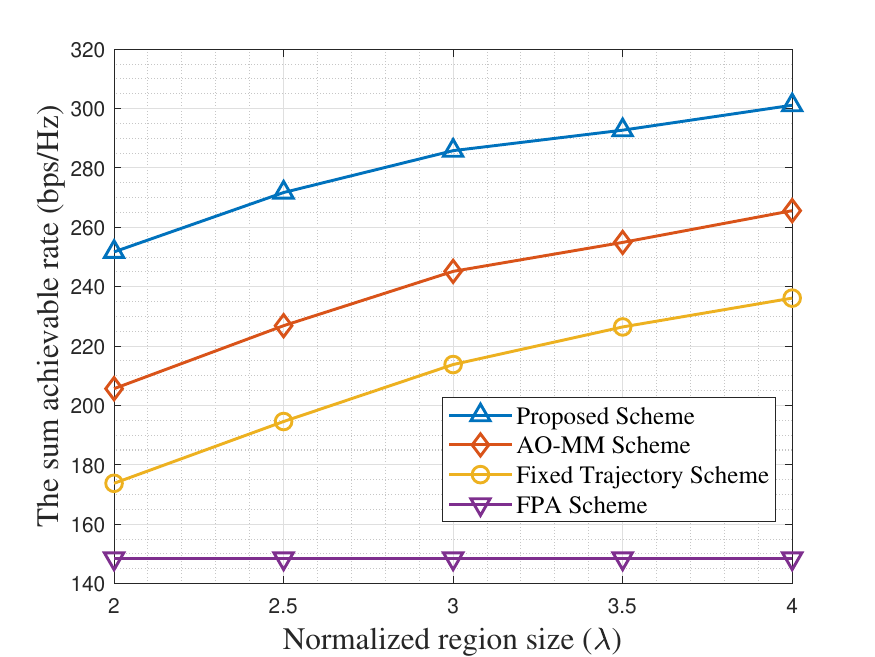}
	\caption{The sum achievable rate versus the normalized region size.}
	\label{antenna_size}
\end{figure}
Fig. \ref{antenna_size} shows the sum achievable rate versus the normalized region size.
As the normalized region size expands, the sum achievable rate of the schemes with antenna position optimization, i.e., the proposed scheme, the AO-MM scheme, and the fixed trajectory scheme, are increased, while the FPA scheme remains static due to immutable antenna positions.
We can observe that the sum achievable rate of the AO-MM scheme is lower than the proposed scheme, which can be attributed to the poorer ability to optimize the antenna position.
The fixed trajectory scheme achieves lower sum achievable rate compared to the proposed scheme and the AO-MM scheme since fixed AAV trajectory cannot exploit the mission regions to approach distant users.

\begin{figure}[!htbp]
	\centering
	\includegraphics[width=0.88\linewidth]{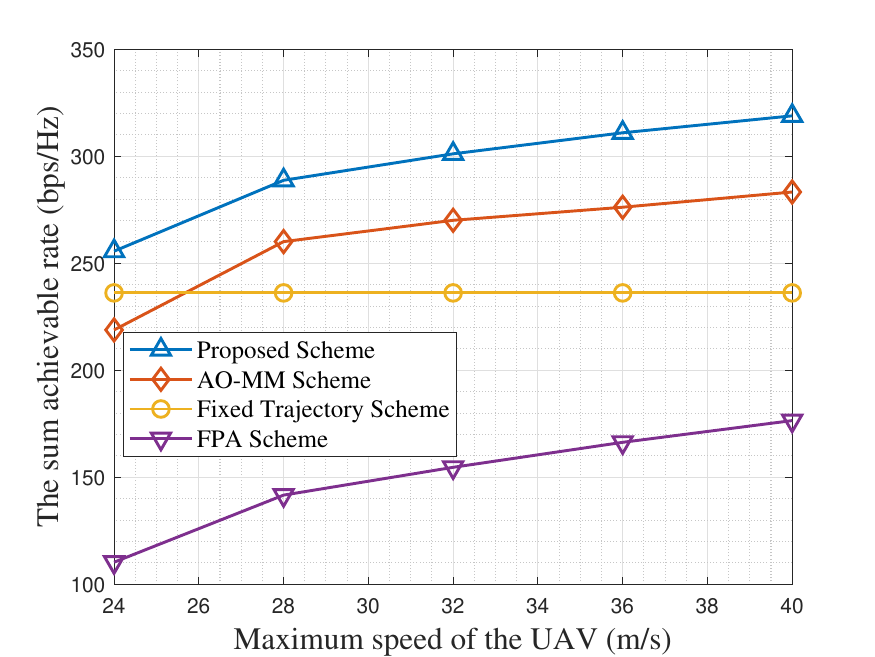}
	\caption{The sum achievable rate versus the AAV maximum speed.}
	\label{AAV_speed}
\end{figure}
Fig. \ref{AAV_speed} presents the sum achievable rate versus the AAV maximum speed, which reveals a fundamental trade-off between the AAV mobility and communication performance.
As the maximum AAV speed increases from 24 $\mathrm{m}/\mathrm{s}$ to 40 $\mathrm{m}/\mathrm{s}$, all trajectory-optimized schemes, i.e., the proposed scheme, the AO-MM scheme, and the FPA scheme, exhibit significant sum achievable rate improvements.
The reason is that with increasing maximum speed of the AAV, the propagation distances between the AAV to certain users can be reduced, thus improving the SINR of the certain users.
The proposed scheme achieves superior sum achievable rate across all schemes, due to the joint optimization of the AAV trajectory, the antenna positions, and the receive beamforming during movement.
Notably, the fixed trajectory scheme shows negligible improvement despite speed increases.
The reason is that, its static trajectory fails to exploit mobility for distance reduction, resulting in wasteful energy expenditure without rate benefits. This underscores the necessity of integrated mobility-aware optimization for energy-efficient AAV communication systems.

\begin{figure}[!htbp]
	\centering
	\includegraphics[width=0.88\linewidth]{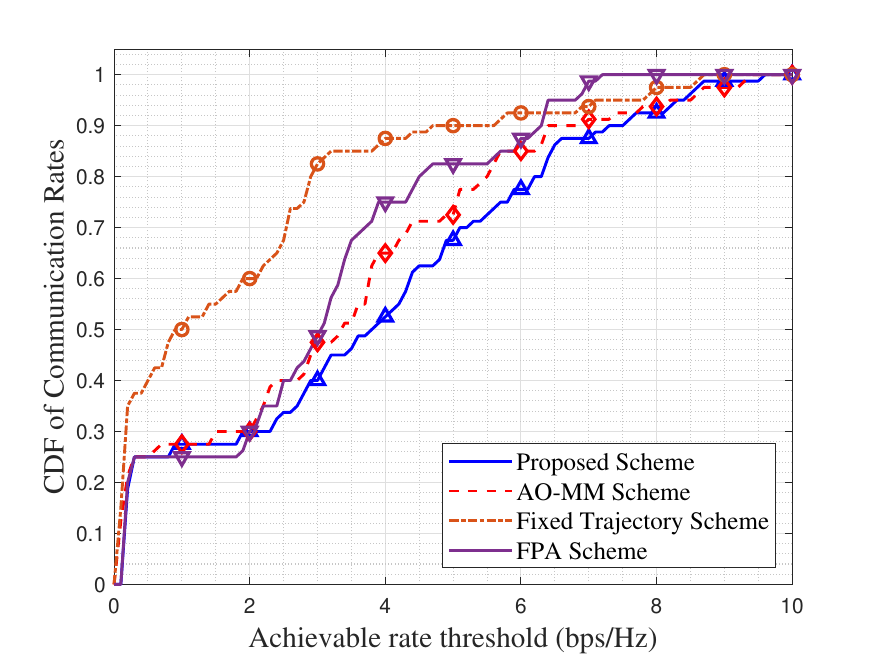}
	\caption{The CDFs of communication rates versus the achievable rate threshold.}
	\label{CDF}
\end{figure}
Fig. \ref{CDF} presents the cumulative distribution functions (CDFs) of communication rates versus the achievable rate threshold, in the perspective of service reliability.
We can observe that, 90\% of users of the proposed scheme achieves over $7.5$ $\mathrm{bps}/\mathrm{Hz}$, reflecting robust interference management and consistent channel quality across the mission period.
Most critically, the fixed trajectory scheme and the FPA scheme show severe user QoS degradation, e.g., 30\% of users in the fixed trajectory scheme fall below 0.2 $\mathrm{bps}/\mathrm{Hz}$ due to static trajectory, resulting in beam-user misalignment.

\section{Conclusion}
In this paper, we investigate an MA-empowered AAV-enabled uplink transmission system, where an AAV is dispatched to collect data from ground users within a designated mission area.
We formulate an uplink sum achievable rate maximization problem of all ground users, by jointly optimizing the AAV trajectory, the AAV receive beamforming, and the antenna positions of the MAs.
We develop an efficient AO-based algorithm to solve the problem, and analyze the computational complexity and convergence of the proposed scheme to evaluate the executability.
The extensive numerical results demonstrate the superior sum achievable rate gains and the service reliability of the proposed scheme compared with several benchmark schemes.

\begin{appendices}
\section{Proof of Theorem \ref{Theorem1}}\label{appendice1}
\begin{proof} The Lagrangian function of problem $(\textbf{P3-2})$ related to receive beamforming can be expressed as
\begin{equation}
    \begin{split}
\mathbf{\mathcal{L}} \left( \Theta_{1}  \right)  = 
\sum_{n=1}^{N} \sum_{m=1}^{M}  \Tilde{R}_{m,n} - \sum_{n=1}^{N} \sum_{m=1}^{M}  \lambda_{m,n} \left( \Vert \mathbf{w}_{m,n} \Vert^2 - 1 \right),
    \end{split}
\end{equation}
where $\Theta_{1} = \{ \mathbf{w}_{m,n}, \lambda_{m,n} \}$. $\{\lambda_{m,n},\forall m,\forall n\}$ are the dual variables associated with the corresponding constraints (\ref{p1d}).
Thus, the derivations of $\mathbf{\mathcal{L}} \left( \Theta_{1}  \right)$ with respect to $\mathbf{w}_{m,n}$ can be given by 
\begin{equation}
    \begin{split}
&\frac{\partial \mathbf{\mathcal{L}} \left( \Theta_{1}  \right)}{\partial \mathbf{w}_{m,n}} 
= - \lambda_{m,n} \mathbf{w}_{m,n}^{*} +  \omega_{m,n} \sqrt{p_{m,n}} \mathbf{h}_{m,n}^{*} \beta_{m,n} \\
& - \omega_{m,n} \left| \beta_{m,n} \right|^{2} \left( 
\sum\limits_{r=1}^{ \mathcal{M}} p_{r,n} \mathbf{h}_{m,n}^{*} \mathbf{h}_{m,n}^{\mathsf{T}} + \sigma_{\mathrm{AAV}}^{2} \mathbf{I}  \right) \mathbf{w}_{m,n}^{*}.
    \end{split}
\end{equation}

By applying the first-order optimality condition, the optimal solution to problem $(\textbf{P3-2})$ can be obtained by setting the derivatives of $\mathbf{\mathcal{L}} \left( \Theta_{1} \right)$ to zero. Thus, the optimal receive beamforming vectors can be given by
\begin{equation}
    \begin{split}
&\mathbf{w}_{m,n}^{\rm opt} = \\
&\frac{ \omega_{m,n} \sqrt{p_{m,n}} \mathbf{h}_{m,n} \beta_{m,n}^{*}  }{ \omega_{m,n} \left| \beta_{m,n} \right|^{2} \left( 
\sum\limits_{r=1}^{ \mathcal{M}} p_{r,n} \mathbf{h}_{m,n}^{*} \mathbf{h}_{m,n}^{\mathsf{T}} + \sigma_{\mathrm{AAV}}^{2} \mathbf{I}  \right) + \lambda_{m,n} \mathbf{I}  }.
    \end{split}
\end{equation}
\end{proof}

\section{Proof of Theorem \ref{Theorem2}}\label{appendice2}
\begin{proof} The Lagrangian function of problem $(\textbf{P3-3})$ related to user transmit power control can be expressed as
\begin{equation}
    \begin{split}
\mathbf{\mathcal{L}} \left( \Theta_{2}  \right)  = 
\sum_{n=1}^{N} \sum_{m=1}^{M}  \Tilde{R}^{\rm{new}}_{m,n} - \sum_{n=1}^{N} \sum_{m=1}^{M}  \mu_{m,n} \left( p_{m,n}^{\rm{new}} - \sqrt{P_{\rm{max}}} \right),
    \end{split}
\end{equation}
where $\Theta_{2} = \{ p_{m,n}^{\rm{new}}, \mu_{m,n} \}$. $\{\mu_{m,n},\forall m,\forall n\}$ are the dual variables associated with the corresponding constraint (\ref{p3-3b}).
Thus, the derivations of $\mathbf{\mathcal{L}} \left( \Theta_{2}  \right)$ with respect to $p_{m,n}^{\rm{new}}$ can be given by 
\begin{equation}
    \begin{split}
\frac{\partial \mathbf{\mathcal{L}} \left( \Theta_{2}  \right)}{\partial p_{m,n}^{\rm{new}} } 
=& - \mu_{m,n}  +  2 \omega_{m,n} {\rm Re} \{ \beta_{m,n}^{*}   \mathbf{w}_{m,n}^{\mathsf{H}} \mathbf{h}_{m,n} \} \\
& - \sum_{r=1}^{M} \omega_{r,n} \left| \beta_{r,n} \right|^{2} \left( 2 p_{m,n}^{\rm{new}} \left|  \mathbf{w}_{r,n}^{\mathsf{H}} \mathbf{h}_{m,n}  \right|^{2}  \right) .
    \end{split}
\end{equation}

By applying the first-order optimality condition, the optimal solution to problem $(\textbf{P3-3})$ can be obtained by setting the derivatives of $\mathbf{\mathcal{L}} \left( \Theta_{2} \right)$ to zero. Thus, the optimal receive beamforming vectors can be given by
\begin{equation}
    \begin{split}
(p_{m,n}^{\rm{new}})^{\rm opt} = \frac{ -\mu_{m,n} + 2 \omega_{m,n} {\rm Re} \{ \beta_{m,n}^{*}   \mathbf{w}_{m,n}^{\mathsf{H}} \mathbf{h}_{m,n} \}  }{ 2 \sum_{r=1}^{M} \omega_{r,n} \left| \beta_{r,n} \right|^{2}  \left|  \mathbf{w}_{r,n}^{\mathsf{H}} \mathbf{h}_{m,n}  \right|^{2}  }.
    \end{split}
\end{equation}
\end{proof}

\section{Proof of the Proposition \ref{pconvergence}}
\label{propostionconv}
\begin{proof}
Let ${\bf{Q}}^{i}$,$ {\bf{W}}^{i}$,$ {\bf{P}}^{i}$, and $ {\bf{U}}^{i}$ denote the solution in the $i$-th iteration of Algorithm \ref{algoverall}, and $\Bar{\Bar{\mathcal{R}}}\left({\bf{Q}}^{i}, {\bf{W}}^{i}, {\bf{P}}^{i}, {\bf{U}}^{i}\right)$ denote the objective function of problem (\textbf{P1}).

Given $ {\bf{W}}^{i}$,$ {\bf{P}}^{i}$, and $ {\bf{U}}^{i}$, we solve problem (\textbf{P2-1}) via Algorithm \ref{Algtra} to get the solution ${\bf{Q}}^{i+1}$. Thus, we have
\begin{equation}
     \Bar{\Bar{\mathcal{R}}}\left({\bf{Q}}^{i+1}, {\bf{W}}^{i}, {\bf{P}}^{i}, {\bf{U}}^{i}\right)
\geq \Bar{\Bar{\mathcal{R}}}\left({\bf{Q}}^{i}, {\bf{W}}^{i}, {\bf{P}}^{i}, {\bf{U}}^{i}\right).
\label{pps1}
\end{equation}
Then, given $ {\bf{Q}}^{i+1}$ and $ {\bf{U}}^{i}$, we solve problem (\textbf{P3-1}) via Algorithm \ref{AlgBCA} 
to get the AAV receive beamforming $\mathbf{W}^{i+1}$, and user transmit power $\mathbf{P}^{i+1}$.
Attribute to the BCA, we have
\begin{equation}
     \Bar{\Bar{\mathcal{R}}}\left({\bf{Q}}^{i+1}, {\bf{W}}^{i+1}, {\bf{P}}^{i+1}, {\bf{U}}^{i}\right)
\geq \Bar{\Bar{\mathcal{R}}}\left({\bf{Q}}^{i+1}, {\bf{W}}^{i}, {\bf{P}}^{i}, {\bf{U}}^{i}\right).
\label{pps2}
\end{equation}
Then, we solve problem (\textbf{P4-1}) via Algorithm \ref{algorithmPSO} to get the antenna position $\mathbf{U}^{i+1}$ for given $ {\bf{Q}}^{i+1}$,$ {\bf{W}}^{i+1}$, and $ {\bf{P}}^{i+1}$.
Since the fitness value is non-decreasing, we have
\begin{equation}
     \Bar{\Bar{\mathcal{R}}}\left({\bf{Q}}^{i+1}, {\bf{W}}^{i+1}, {\bf{P}}^{i+1}, {\bf{U}}^{i+1}\right)
\geq \Bar{\Bar{\mathcal{R}}}\left({\bf{Q}}^{i+1}, {\bf{W}}^{i+1}, {\bf{P}}^{i+1}, {\bf{U}}^{i}\right).
\label{pps3}
\end{equation}
Combining \eqref{pps1}, \eqref{pps2}, and \eqref{pps3}, we have
\begin{equation}
     \Bar{\Bar{\mathcal{R}}}\left({\bf{Q}}^{i+1}, {\bf{W}}^{i+1}, {\bf{P}}^{i+1}, {\bf{U}}^{i+1}\right)
\geq \Bar{\Bar{\mathcal{R}}}\left({\bf{Q}}^{i}, {\bf{W}}^{i}, {\bf{P}}^{i}, {\bf{U}}^{i}\right),
\end{equation}
which indicates that the value of the objective function \eqref{p1o} is non-decreasing over iterations. Moreover, the sum-rate for all users is constrained by the resource constraints on communication, AAV mobility, and energy.
Hence, Algorithm \ref{algoverall} is guaranteed to converge.
\end{proof}

\section{Antenna Position Optimization via MM}
\label{AOMM}
In this appendix, we present a benchmark scheme based on AO, where the MA positions are optimized via the MM method while other variables follow identical optimization procedures to our proposed scheme.

Specifically, the field response channel matrix $\mathbf{H}_{m,n}$ exhibits a complex and nonlinear relationship with respect to the MA position $\mathbf{U}$. Accordingly, we transform the problem using the WMMSE method, and the optimization problem can be reformulated as
\begin{subequations} 
\begin{flalign}
 (\textbf{P\ref{AOMM}-1}):\ & \max_{\mathbf{U},\boldsymbol{\beta}, \boldsymbol{\omega}} \quad \sum_{n=1}^{N} \sum_{m=1}^{M}  \Tilde{R}_{m,n} \\
 {\rm{s.t.}}  \quad &\eqref{p1e}-\eqref{p1f},\\
 & \omega_{m,n}\geq 0, \forall m, \forall n.
\end{flalign}
\end{subequations}
Then, we optimize the auxiliary variables based on Eqs. \eqref{p3beta} and \eqref{p3omega}.
Similar to pioneer works \cite{Feng2024Weighted, Ma2024MIMO}, to reduce computational complexity, we optimize each MA position separately within each time slot, thereby decoupling the relationships between the MAs.
Specifically, when optimizing $\{ \mathbf{u}_{k,n}, \forall n \}$, all other variables $\{ \mathbf{u}_{k',n}, k' \neq k, \forall n \}$ are held fixed. By doing so, the original optimization problem can be decomposed into $K$ separate subproblems, each addressing the optimization of a single MA position.
\begin{subequations} 
\begin{flalign}
 (\textbf{P\ref{AOMM}-1}.k):\ & \max_{\mathbf{u}_{k,n},\forall n} \quad \sum_{n=1}^{N} \sum_{m=1}^{M}  \Tilde{R}_{m,n} \\
 {\rm{s.t.}}  \quad  & \| \mathbf{u}_{k,n} -\mathbf{u}_{k',n}\| \ge d_{\min}, \forall k \neq k', \forall k \in \mathcal{K}, \forall n \in \mathcal{N}, \label{p4-1a}  \\
 & \mathbf{u}_{k,n} \in \mathcal{R}_{s}, \forall k\in \mathcal{K}, \forall n\in \mathcal{N}. \label{p4-1b}
\end{flalign}
\end{subequations}

Regarding $\Tilde{R}_{m,n}$, it can be expressed as a function of $\mathbf{u}_{k,n}$, as shown below.
\begin{equation}
\begin{split}
\Tilde{R}_{m,n} =& - \omega_{m,n} \left| \beta_{m,n} \right|^{2} \sum\limits_{r=1}^{M} p_{r,n} \left|  \mathbf{w}_{m,n}^{\mathsf{H}} \mathbf{h}_{r,n}  \right|^{2} \\
& + 2 \omega_{m,n} {\rm Re} \{ \beta_{m,n}^{*} \sqrt{p_{m,n}}  \mathbf{w}_{m,n}^{\mathsf{H}} \mathbf{h}_{m,n} \} + \Upsilon_{m,n},
\end{split}
\end{equation}
where 
\begin{equation}
\Upsilon_{m,n} = \log_{2}(\omega_{m,n}) - \omega_{m,n} - \omega_{m,n}\left| \beta_{m,n} \right|^{2}\| \mathbf{w}_{m,n}^{\mathsf{H}} \|^{2}_{2}
\sigma_{\mathrm{AAV}}^{2} +1.
\end{equation}
For notational convenience, we denote the entrie in the $i$-th row of $\mathbf{w}_{m,n}$ as $w_{m,n,i}$.
Then, based on (\ref{channel_vector}), it follows that 
\begin{equation}
\begin{split}
\left| \mathbf{w}_{m,n}^{\mathsf{H}} \mathbf{h}_{r,n}  \right|^{2} =&
g_{r,k,n}^{\mathsf{H}} \mathbf{A}_{m,r,k,n} g_{r,k,n} + {\rm Re}\{ \mathbf{b}_{m,r,k,n}^{\mathsf{H}} g_{r,k,n} \} \\
& + \mathbf{c}_{m,r,k,n},
\end{split}
\end{equation}
\begin{equation}
{\rm Re} \{ \mathbf{w}_{m,n}^{\mathsf{H}} \mathbf{h}_{m,n} \} = 
{\rm Re} \{ \sum_{k=1}^{K}w_{m,n,k} \mathbf{f}_{m}^{\mathsf{H}} \mathbf{\Sigma}_{m,n}^{\mathsf{H}} g_{m,k,n} \},
\end{equation}
where
\begin{equation}
    \begin{split}
\mathbf{A}_{m,r,k,n} = w_{m,n,k}w_{m,n,k}^{*} \mathbf{\Sigma}_{r,n} \mathbf{f}_{r} \mathbf{f}_{r}^{\mathsf{H}} \mathbf{\Sigma}_{r,n}^{\mathsf{H}}, 
    \end{split}
\end{equation}
\begin{equation}
    \begin{split}
\mathbf{b}_{m,r,k,n} = 2 \sum_{i=1,i \neq k}^{K} \mathbf{f}_{r}^{\mathsf{H}} \mathbf{\Sigma}_{r,n}^{\mathsf{H}}g_{r,i,n}\mathbf{\Sigma}_{r,n} \mathbf{f}_{r} w_{m,n,i}w_{m,n,k}^{\rm{*}}, 
    \end{split}
\end{equation}
\begin{equation} 
    \begin{split}
\mathbf{c}_{m,r,k,n} =& \left(\sum_{i=1,i \neq k}^{K}w_{m,n,i}\mathbf{f}_{r}^{\mathsf{H}} \mathbf{\Sigma}_{r,n}^{\mathsf{H}}g_{r,i,n}\right)\\
&\times \left(\sum_{j=1,j \neq k}^{K}w_{m,n,j}^{\rm{*}}
g_{r,j,n}^{\mathsf{H}} \mathbf{f}_{r} \mathbf{\Sigma}_{r,n}\right). 
    \end{split}
\end{equation}
Then, $\Tilde{R}_{m,n}$ can be expressed as
\begin{equation}
\begin{split}
\Tilde{R}_{m,n} = & g_{m,k,n}^{\mathsf{H}} \mathbf{E}_{m,k,n} g_{m,k,n} + {\rm Re} \{ \mathbf{F}_{m,k,n}^{\mathsf{H}} g_{m,k,n}  \} +   \Tilde{\Upsilon}_{m,n},
\end{split}
\end{equation}
where 
\begin{equation}
\begin{split}
\mathbf{E}_{m,k,n} = - \omega_{m,n} \left| \beta_{m,n} \right|^{2} \sum\limits_{r=1}^{M} p_{r,n} \mathbf{A}_{m,r,k,n},
\end{split}
\end{equation}
\begin{equation}
\begin{split}
\mathbf{F}_{m,k,n} =& - \omega_{m,n} \left| \beta_{m,n} \right|^{2} \sum\limits_{r=1}^{M} p_{r,n} \mathbf{b}_{m,r,k,n} \\
 & + 2 \omega_{m,n} w_{m,n,k} \mathbf{f}_{m}^{\mathsf{H}} \mathbf{\Sigma}_{m,n}^{\mathsf{H}},
\end{split}
\end{equation}
\begin{equation}
\begin{split}
\Tilde{\Upsilon}_{m,n} =& \Upsilon_{m,n} + {\rm Re} \{ \sum_{i=1,i\neq k}^{K}w_{m,n,i} \mathbf{f}_{m}^{\mathsf{H}} \mathbf{\Sigma}_{m,n}^{\mathsf{H}} g_{m,i,n}  \} \\
&- \omega_{m,n} \left| \beta_{m,n} \right|^{2} \sum\limits_{r=1}^{M} p_{r,n} \mathbf{c}_{m,r,k,n}.
\end{split}
\end{equation}

We apply the MM method to tackle the optimization problem \cite{Feng2024Weighted}. The effectiveness of the MM technique hinges on developing a surrogate function that adheres to the upper bound property for both the objective function and its constraints. In what follows, we present several lemmas that aid in the practical application of the MM method.

\begin{lemma} \label{lemma1}
The quadratic form $\mathbf{x}^{\mathsf{H}} \mathbf{L} \mathbf{x}$, where $\mathbf{L}$ is a Hermitian matrix, can be upper bounded as \cite{Feng2024Weighted, sun2016majorization}:
\begin{equation}
\mathbf{x}^{\mathsf{H}} \mathbf{L} \mathbf{x} \leq \mathbf{x}^{\mathsf{H}} \mathbf{M} \mathbf{x} + 2 \mathrm{Re}(\mathbf{x}^{\mathsf{H}} (\mathbf{L} - \mathbf{M}) \mathbf{x}_0) + \mathbf{x}_0^{\mathsf{H}} (\mathbf{M} - \mathbf{L}) \mathbf{x}_0,
\end{equation}
where $\mathbf{M} \succeq \mathbf{L}$. Equality is achieved at $\mathbf{x} = \mathbf{x}_0$.
\end{lemma}

\begin{lemma} \label{lemma2}
$\Psi_{m,n}(\mathbf{u}_{k,n}) = {\rm Re} \{ \mathbf{\Gamma}^{\mathsf{H}} g_{m,k,n}(\mathbf{u}_{k,n})  \} $ can be bounded from both below and above as follows \cite{Ma2024MIMO}
\begin{equation}
\begin{split}
\Psi_{m,n}(\mathbf{u}_{k,n}) &\geq \Psi_{m,n}(\mathbf{u}_{k,n}^{(l)}) + \nabla \Psi_{m,n}(\mathbf{u}_{k,n}^{(l)})^{\mathsf{T}} (\mathbf{u}_{k,n} - \mathbf{u}_{k,n}^{(l)})  \\
&- \frac{4\pi^2}{\lambda^2} \| \mathbf{\Gamma} \|_1 (\mathbf{u}_{k,n} - \mathbf{u}_{k,n}^{(l)})^{\mathsf{T}} (\mathbf{u}_{k,n} - \mathbf{u}_{k,n}^{(l)}),
\end{split}
\end{equation}
\begin{equation}
\begin{split}
\Psi_{m,n}(\mathbf{u}_{k,n}) &\leq \Psi_{m,n}(\mathbf{u}_{k,n}^{(l)}) + \nabla \Psi_{m,n}(\mathbf{u}_{k,n}^{(l)})^{\mathsf{T}} (\mathbf{u}_{k,n} - \mathbf{u}_{k,n}^{(l)})  \\
&+ \frac{4\pi^2}{\lambda^2} \| \mathbf{\Gamma}\|_1 (\mathbf{u}_{k,n} - \mathbf{u}_{k,n}^{(l)})^{\mathsf{T}} (\mathbf{u}_{k,n} - \mathbf{u}_{k,n}^{(l)}),
\end{split}
\end{equation}
where
\begin{equation}
\nabla \Psi_{m,n}(\mathbf{u}_{k,n}^{(l)}) = {\rm Re} \{ \mathbf{\Gamma}^{\mathsf{H}} \nabla g_{m,k,n}^{(l)}) \}^{\mathsf{T}},
\end{equation}
\begin{equation}\small
\begin{split} 
\nabla g_{m,k,n}^{(l)} 
 = j \frac{2\pi}{\lambda} (\mathbf{n}_{m,1,n}, \cdots, \mathbf{n}_{m,L_{m}^{t},n})^{\mathsf{T}} \odot (g_{m,k,n}^{(l)}, g_{m,k,n}^{(l)}) ),
\end{split}
\end{equation}
\begin{equation}
\mathbf{n}_{m,i,n} = (\sin \theta_{m,i,n}\cos\phi_{m,i,n}, \sin\theta_{m,i,n}\sin\phi_{m,i,n})^{\mathsf{T}},
\end{equation}
where we denote  $g_{m,k,n}(\mathbf{u}_{k,n}^{(l)})$ as $g_{m,k,n}^{(l)}$, for notational convenience.
The equality is achieved at $\mathbf{u}_{k,n} = \mathbf{u}_{k,n}^{(l)}$.

\end{lemma}

Based on Lemmas \ref{lemma1} and \ref{lemma2}, a minorizing function for the objective function can be expressed by \eqref{p4-1-1} at the $l$-th iteration and evaluated at $\mathbf{u}_{k,n} = \mathbf{u}_{k,n}^{(l)}$, where $\mathbf{\Gamma} = \mathbf{J}_{m,k,n}$, and

\begin{figure*}[!t]
\vspace*{-\baselineskip} 
\begin{equation} \small
\begin{aligned} 
\Tilde{R}_{m,n}& =  g_{m,k,n}^{\mathsf{H}} \mathbf{E}_{m,k,n} g_{m,k,n} + {\rm Re} \{ \mathbf{F}_{m,k,n}^{\mathsf{H}} g_{m,k,n}  \} +   \Tilde{\Upsilon}_{m,n} \\
& \ge \zeta_{m,k,n} g_{m,k,n}^{\mathsf{H}} \mathbf{I} g_{m,k,n} + 2 {\rm Re} \{ g_{m,k,n}^{\mathsf{H}} (\mathbf{E}_{m,k,n} - \zeta_{m,k,n}\mathbf{I} ) g_{m,k,n}^{(l)} \} + (g_{m,k,n}^{(l)})^{\mathsf{H}} ( \zeta_{m,k,n}\mathbf{I} -  \mathbf{E}_{m,k,n})g_{m,k,n}^{(l)} \\
&\quad\ +\ {\rm Re} \{ \mathbf{F}_{m,k,n}^{\mathsf{H}} g_{m,k,n}  \} +   \Tilde{\Upsilon}_{m,n} \\
& = {\rm Re} \{ \underbrace{ ( ( 2\mathbf{E}_{m,k,n}g_{m,k,n}^{(l)} - 2\zeta_{m,k,n}\mathbf{I}g_{m,k,n}^{(l)})^{\mathsf{H}}  + \mathbf{F}_{m,k,n}^{\mathsf{H}} ) }_{ \mathbf{J}_{m,k,n}^{\mathsf{H}} } g_{m,k,n} \} + (g_{m,k,n}^{(l)})^{\mathsf{H}} ( \zeta_{m,k,n}\mathbf{I} -  \mathbf{E}_{m,k,n})g_{m,k,n}^{(l)} + \Tilde{\Upsilon}_{m,n} \\
& \ge {\rm Re} \{ \mathbf{J}_{m,k,n}^{\mathsf{H}} g_{m,k,n}^{(l)}  \}\Psi_{m,n}(\mathbf{u}_{k,n}^{(l)}) + \nabla \Psi_{m,n}(\mathbf{u}_{k,n}^{(l)})^{\mathsf{T}} (\mathbf{u}_{k,n} - \mathbf{u}_{k,n}^{(l)}) + \frac{4\pi^2}{\lambda^2} \| \mathbf{J}_{m,k,n} \|_1 (\mathbf{u}_{k,n} - \mathbf{u}_{k,n}^{(l)})^{\mathsf{T}} (\mathbf{u}_{k,n} - \mathbf{u}_{k,n}^{(l)}) \\
& = \Tilde{R}_{m,n}^{\rm{new}},
\label{p4-1-1}
\end{aligned}
\end{equation}
\hrulefill
\end{figure*}

\begin{equation}
\begin{split}
\zeta_{m,k,n} = - \omega_{m,n} \left| \beta_{m,n} \right|^{2} \sum\limits_{r=1}^{M} p_{r,n} w_{m,n,k}w_{m,n,k}^{*} 
\| \mathbf{\Sigma}_{r,n} \mathbf{f}_{r} \|^{2}.
\end{split}
\end{equation}

Based on the Cauchy-Schwarz inequality, a minorizing function for the constraint is developed at $\mathbf{u}_{k,n} = \mathbf{u}_{k,n}^{(l)}$ during the $l$-th iteration, as described below
\begin{equation}
    \| \mathbf{u}_{k,n} - \mathbf{u}_{k',n} \|_2 
    \geq 
    \frac{(\mathbf{u}_{k,n}^{(l)} - \mathbf{u}_{k',n})^{\mathsf{T}} (\mathbf{u}_{k,n} - \mathbf{u}_{k',n})}{\| \mathbf{u}_{k,n}^{(l)} - \mathbf{u}_{k',n} \|_2}. \label{p4-1-2}
\end{equation}

In the end, the MA position optimization problem can be expressed as
\begin{subequations} 
\begin{flalign}
 (\textbf{P\ref{AOMM}-2}.k):\ & \max_{\mathbf{u}_{k,n},\forall n} \quad \sum_{n=1}^{N} \sum_{m=1}^{M}  \Tilde{R}_{m,n}^{\rm{new}} \\
 {\rm{s.t.}}  \quad  & \eqref{p4-1b}, \eqref{p4-1-2}.
\end{flalign}
\end{subequations}
This is a typical convex quadratic programming (QP) problem and can be efficiently solved using quadprog or CVX.

\end{appendices}

\bibliographystyle{IEEEtran}
\bibliography{myref}

\begin{thebibliography}{10}
\providecommand{\url}[1]{#1}
\csname url@samestyle\endcsname
\providecommand{\newblock}{\relax}
\providecommand{\bibinfo}[2]{#2}
\providecommand{\BIBentrySTDinterwordspacing}{\spaceskip=0pt\relax}
\providecommand{\BIBentryALTinterwordstretchfactor}{4}
\providecommand{\BIBentryALTinterwordspacing}{\spaceskip=\fontdimen2\font plus
\BIBentryALTinterwordstretchfactor\fontdimen3\font minus \fontdimen4\font\relax}
\providecommand{\BIBforeignlanguage}[2]{{%
\expandafter\ifx\csname l@#1\endcsname\relax
\typeout{** WARNING: IEEEtran.bst: No hyphenation pattern has been}%
\typeout{** loaded for the language `#1'. Using the pattern for}%
\typeout{** the default language instead.}%
\else
\language=\csname l@#1\endcsname
\fi
#2}}
\providecommand{\BIBdecl}{\relax}
\BIBdecl

\bibitem{10879807}
G.~Cheng, X.~Song, Z.~Lyu, and J.~Xu, ``Networked {ISAC} for low-altitude economy: Coordinated transmit beamforming and {UAV} trajectory design,'' \emph{IEEE Trans. Commun.}, to appear, 2025.

\bibitem{10693833}
Z.~Kuang, W.~Liu, C.~Wang, Z.~Jin, J.~Ren, X.~Zhang, and Y.~Shen, ``Movable-antenna array empowered {ISAC} systems for low-altitude economy,'' in \emph{Proc. IEEE/CIC Int. Conf. Commun. China Workshops (ICCC Workshops)}, Hangzhou, China, Aug. 2024, pp. 776--781.

\bibitem{9456851}
Q.~Wu, J.~Xu, Y.~Zeng, D.~W.~K. Ng, N.~Al-Dhahir, R.~Schober, and A.~L. Swindlehurst, ``A comprehensive overview on {5G}-and-beyond networks with {UAVs}: From communications to sensing and intelligence,'' \emph{IEEE J. Sel. Areas Commun.}, vol.~39, no.~10, pp. 2912--2945, Oct. 2021.

\bibitem{8918497}
Y.~Zeng, Q.~Wu, and R.~Zhang, ``Accessing from the sky: A tutorial on {UAV} communications for {5G} and beyond,'' \emph{Proc. IEEE}, vol. 107, no.~12, pp. 2327--2375, Dec. 2019.

\bibitem{10606316}
W.~Liu, H.~Wang, X.~Zhang, H.~Xing, J.~Ren, Y.~Shen, and S.~Cui, ``Joint trajectory design and resource allocation in {UAV}-enabled heterogeneous {MEC} systems,'' \emph{IEEE Internet Things J.}, vol.~11, no.~19, pp. 30\,817--30\,832, Oct. 2024.

\bibitem{7894280}
M.~Shafi, A.~F. Molisch, P.~J. Smith, T.~Haustein, P.~Zhu, P.~De~Silva, F.~Tufvesson, A.~Benjebbour, and G.~Wunder, ``{5G}: A tutorial overview of standards, trials, challenges, deployment, and practice,'' \emph{IEEE J. Sel. Areas Commun.}, vol.~35, no.~6, pp. 1201--1221, Jun. 2017.

\bibitem{8030501}
A.~F. Molisch, V.~V. Ratnam, S.~Han, Z.~Li, S.~L.~H. Nguyen, L.~Li, and K.~Haneda, ``Hybrid beamforming for massive {MIMO}: A survey,'' \emph{IEEE Commun. Mag.}, vol.~55, no.~9, pp. 134--141, Sep. 2017.

\bibitem{9461747}
W.~Feng, J.~Tang, N.~Zhao, X.~Zhang, X.~Wang, K.-K. Wong, and J.~A. Chambers, ``Hybrid beamforming design and resource allocation for {UAV}-aided wireless-powered mobile edge computing networks with {NOMA},'' \emph{IEEE J. Sel. Areas Commun.}, vol.~39, no.~11, pp. 3271--3286, Nov. 2021.

\bibitem{10972180}
Z.~Li, J.~Ba, Z.~Su, J.~Huang, H.~Peng, W.~Chen, L.~Du, and T.~H. Luan, ``Movable antennas enabled {ISAC} systems: Fundamentals, opportunities, and future directions,'' \emph{IEEE Wireless Commun.}, to appear, 2025.

\bibitem{zhu2024historical}
L.~Zhu and K.-K. Wong, ``Historical review of fluid antenna and movable antenna,'' \emph{arXiv preprint arXiv:2401.02362}, 2024.

\bibitem{10286328}
L.~Zhu, W.~Ma, and R.~Zhang, ``Movable antennas for wireless communication: Opportunities and challenges,'' \emph{IEEE Commun. Mag.}, vol.~62, no.~6, pp. 114--120, Jun. 2024.

\bibitem{liu2025movable}
W.~Liu, X.~Zhang, C.~Wang, J.~Ren, and W.~Yuan, ``Movable antennas meet low-altitude wireless networks: Fundamentals, opportunities, and future directions,'' \emph{arXiv preprint arXiv:2506.13250}, 2025.

\bibitem{10318061}
L.~Zhu, W.~Ma, and R.~Zhang, ``Modeling and performance analysis for movable antenna enabled wireless communications,'' \emph{IEEE Trans. Wireless Commun.}, vol.~23, no.~6, pp. 6234--6250, Jun. 2024.

\bibitem{11017619}
C.~Dou, Y.~Wu, L.~Qian, K.-K. Wong, and T.~Q.~S. Quek, ``Fluid antenna empowered integration of sensing, communications and computing with hybrid multi-task offloading,'' \emph{IEEE Wireless Commun. Lett.}, to appear, 2025.

\bibitem{bai2024movable}
Y.~Bai, B.~Xie, R.~Zhu, Z.~Chang, and R.~Jantti, ``Movable antenna-equipped {UAV} for data collection in backscatter sensor networks: A deep reinforcement learning-based approach,'' \emph{arXiv preprint arXiv:2411.13970}, 2024.

\bibitem{8807386}
J.~Cui, Y.~Liu, and A.~Nallanathan, ``Multi-agent reinforcement learning-based resource allocation for {UAV} networks,'' \emph{IEEE Trans. Wireless Commun.}, vol.~19, no.~2, pp. 729--743, Feb. 2020.

\bibitem{8676325}
C.~H. Liu, X.~Ma, X.~Gao, and J.~Tang, ``Distributed energy-efficient multi-{UAV} navigation for long-term communication coverage by deep reinforcement learning,'' \emph{IEEE Trans. Mobile Comput.}, vol.~19, no.~6, pp. 1274--1285, Jun. 2020.

\bibitem{8626132}
F.~Cheng, G.~Gui, N.~Zhao, Y.~Chen, J.~Tang, and H.~Sari, ``{UAV}-relaying-assisted secure transmission with caching,'' \emph{IEEE Trans. Commun.}, vol.~67, no.~5, pp. 3140--3153, May 2019.

\bibitem{9206550}
A.~Ranjha and G.~Kaddoum, ``{URLLC} facilitated by mobile {UAV} relay and {RIS}: A joint design of passive beamforming, blocklength, and {UAV} positioning,'' \emph{IEEE Internet Things J.}, vol.~8, no.~6, pp. 4618--4627, Mar. 2021.

\bibitem{9273074}
W.~Luo, Y.~Shen, B.~Yang, S.~Wang, and X.~Guan, ``Joint 3-{D} trajectory and resource optimization in multi-{UAV}-enabled {IoT} networks with wireless power transfer,'' \emph{IEEE Internet Things J.}, vol.~8, no.~10, pp. 7833--7848, May 2021.

\bibitem{10087216}
S.~Gong, M.~Wang, B.~Gu, W.~Zhang, D.~T. Hoang, and D.~Niyato, ``Bayesian optimization enhanced deep reinforcement learning for trajectory planning and network formation in multi-{UAV} networks,'' \emph{IEEE Trans. Veh. Technol.}, vol.~72, no.~8, pp. 10\,933--10\,948, Aug. 2023.

\bibitem{9916163}
Z.~Lyu, G.~Zhu, and J.~Xu, ``Joint maneuver and beamforming design for {UAV}-enabled integrated sensing and communication,'' \emph{IEEE Trans. Wireless Commun.}, vol.~22, no.~4, pp. 2424--2440, Apr. 2023.

\bibitem{10233771}
W.~Liu, Z.~Jin, X.~Zhang, W.~Zang, S.~Wang, and Y.~Shen, ``{AoI}-aware {UAV}-enabled marine {MEC} networks with integrated sensing, computation, and communication,'' in \emph{Proc. IEEE/CIC Int. Conf. Commun. China Workshops (ICCC Workshops)}, Dalian, China, Aug. 2023, pp. 1--6.

\bibitem{10146439}
S.~Bi, J.~Yu, Z.~Yang, X.~Lin, and Y.~Wu, ``Joint 3-{D} deployment and resource allocation for {UAV}-assisted integrated communication and localization,'' \emph{IEEE Wireless Commun. Lett.}, vol.~12, no.~10, pp. 1672--1676, Oct. 2023.

\bibitem{9822386}
T.~V. Nguyen, H.~D. Le, and A.~T. Pham, ``On the design of {RIS}–{UAV} relay-assisted hybrid {FSO}/{RF} satellite–aerial–ground integrated network,'' \emph{IEEE Trans. Aerosp. Electron. Syst.}, vol.~59, no.~2, pp. 757--771, Apr. 2023.

\bibitem{9904508}
Q.~Xu, Z.~Su, D.~Fang, and Y.~Wu, ``Hierarchical bandwidth allocation for social community-oriented multicast in space-air-ground integrated networks,'' \emph{IEEE Trans. Wireless Commun.}, vol.~22, no.~3, pp. 1915--1930, Mar. 2023.

\bibitem{10233456}
W.~Liu, J.~Wang, H.~Xing, Z.~Jin, X.~Zhang, and Y.~Shen, ``Blockchain-empowered space-air-ground integrated networks for remote internet of things,'' in \emph{Proc. IEEE/CIC Int. Conf. Commun. China}, Dalian, China, Aug. 2023, pp. 1--6.

\bibitem{10271264}
X.~Chen, Z.~Chang, M.~Liu, N.~Zhao, T.~Hämäläinen, and D.~Niyato, ``{UAV}-{IRS} assisted covert communication: Introducing uncertainty via phase shifting,'' \emph{IEEE Wireless Commun. Lett.}, vol.~13, no.~1, pp. 103--107, Jan. 2024.

\bibitem{9903838}
L.~P. Qian, W.~Zhang, Q.~Wang, Y.~Wu, and X.~Yang, ``Alternative optimization for secrecy throughput maximization in {UAV}-aided {NOMA} networks,'' \emph{IEEE Wireless Commun. Lett.}, vol.~11, no.~12, pp. 2580--2584, Dec. 2022.

\bibitem{9604506}
X.~Zhang, W.~Luo, Y.~Shen, and S.~Wang, ``Average {AoI} minimization in {UAV}-assisted {IoT} backscatter communication systems with updated information,'' in \emph{Proc. IEEE Ubiquitous Intell. Comput. (UIC)}, Atlanta, GA, USA, Oct. 2021, pp. 123--130.

\bibitem{10539623}
Y.~Long, S.~Zhao, S.~Gong, B.~Gu, D.~Niyato, and X.~Shen, ``{AoI}-aware sensing scheduling and trajectory optimization for multi-{UAV}-assisted wireless backscatter networks,'' \emph{IEEE Trans. Veh. Technol.}, vol.~73, no.~10, pp. 15\,440--15\,455, Oct. 2024.

\bibitem{9779853}
Z.~Wei, M.~Zhu, N.~Zhang, L.~Wang, Y.~Zou, Z.~Meng, H.~Wu, and Z.~Feng, ``{UAV}-assisted data collection for internet of things: A survey,'' \emph{IEEE Internet Things J.}, vol.~9, no.~17, pp. 15\,460--15\,483, Sep. 2022.

\bibitem{10980172}
X.~Zhang, H.~Xing, Y.~Shen, J.~Xu, and S.~Cui, ``Age of information minimization in {UAV}-enabled {IoT} networks via federated reinforcement learning,'' \emph{IEEE Trans. Wireless Commun.}, to appear, 2025.

\bibitem{9321340}
S.~Fu, Y.~Tang, Y.~Wu, N.~Zhang, H.~Gu, C.~Chen, and M.~Liu, ``Energy-efficient {UAV}-enabled data collection via wireless charging: A reinforcement learning approach,'' \emph{IEEE Internet Things J.}, vol.~8, no.~12, pp. 10\,209--10\,219, Jun. 2021.

\bibitem{8956055}
Z.~Yu, Y.~Gong, S.~Gong, and Y.~Guo, ``Joint task offloading and resource allocation in {UAV}-enabled mobile edge computing,'' \emph{IEEE Internet Things J.}, vol.~7, no.~4, pp. 3147--3159, Apr. 2020.

\bibitem{10100681}
M.~Dai, Z.~Luo, Y.~Wu, L.~Qian, B.~Lin, and Z.~Su, ``Incentive oriented two-tier task offloading scheme in marine edge computing networks: A hybrid stackelberg-auction game approach,'' \emph{IEEE Trans. Wireless Commun.}, vol.~22, no.~12, pp. 8603--8619, Dec. 2023.

\bibitem{9985993}
M.~Fu, Y.~Zhou, Y.~Shi, C.~Jiang, and W.~Zhang, ``{UAV}-assisted multi-cluster over-the-air computation,'' \emph{IEEE Trans. Wireless Commun.}, vol.~22, no.~7, pp. 4668--4682, Jul. 2023.

\bibitem{10946250}
Y.~Chen, S.~Sun, M.~Liu, B.~Ai, Y.~Wang, and Y.~Liu, ``Energy-efficient over-the-air computation in {UAV}-assisted {IIoT} networks,'' \emph{IEEE Trans. Mobile Comput.}, to appear, 2025.

\bibitem{10220154}
M.~Fu, Y.~Shi, and Y.~Zhou, ``Federated learning via unmanned aerial vehicle,'' \emph{IEEE Trans. Wireless Commun.}, vol.~23, no.~4, pp. 2884--2900, Apr. 2024.

\bibitem{10818523}
Z.~Zhai, X.~Yuan, X.~Wang, and H.~Yang, ``{UAV}-enabled asynchronous federated learning,'' \emph{IEEE Trans. Wireless Commun.}, vol.~24, no.~3, pp. 2358--2372, Mar. 2025.

\bibitem{10972043}
X.~Zhang, W.~Liu, J.~Ren, H.~Xing, G.~Gui, Y.~Shen, and S.~Cui, ``Latency minimization for {UAV}-enabled federated learning: Trajectory design and resource allocation,'' \emph{IEEE Internet Things J.}, vol.~12, no.~14, pp. 27\,097--27\,112, Jul. 2025.

\bibitem{10654366}
W.~Liu, X.~Zhang, H.~Xing, J.~Ren, Y.~Shen, and S.~Cui, ``{UAV}-enabled wireless networks with movable-antenna array: Flexible beamforming and trajectory design,'' \emph{IEEE Wireless Commun. Lett.}, vol.~14, no.~3, pp. 566--570, Mar. 2025.

\bibitem{10918750}
T.~Ren, X.~Zhang, L.~Zhu, W.~Ma, X.~Gao, and R.~Zhang, ``{6D} movable antenna enhanced interference mitigation for cellular-connected {UAV} communications,'' \emph{IEEE Wireless Commun. Lett.}, vol.~14, no.~6, pp. 1618--1622, Jun. 2025.

\bibitem{Xu2020Multiuser}
D.~Xu, Y.~Sun, D.~W.~K. Ng, and R.~Schober, ``Multiuser {MISO} {UAV} communications in uncertain environments with no-fly zones: Robust trajectory and resource allocation design,'' \emph{IEEE Trans. Commun.}, vol.~68, no.~5, pp. 3153--3172, May 2020.

\bibitem{Qin2024Antenna}
H.~Qin, W.~Chen, Z.~Li, Q.~Wu, N.~Cheng, and F.~Chen, ``Antenna positioning and beamforming design for fluid antenna-assisted multi-user downlink communications,'' \emph{IEEE Wireless Commun. Lett.}, vol.~13, no.~4, pp. 1073--1077, Apr. 2024.

\bibitem{10901248}
A.~Khalili and R.~Schober, ``Advanced {ISAC} design: Movable antennas and accounting for dynamic {RCS},'' in \emph{Proc. IEEE Glob. Commun. Conf. (GLOBECOM)}, Cape Town, South Africa, Dec. 2024, pp. 4022--4027.

\bibitem{Deng2023Beamforming}
C.~Deng, X.~Fang, and X.~Wang, ``Beamforming design and trajectory optimization for {UAV}-empowered adaptable integrated sensing and communication,'' \emph{IEEE Trans. Wireless Commun.}, vol.~22, no.~11, pp. 8512--8526, Nov. 2023.

\bibitem{Shi2011An}
Q.~Shi, M.~Razaviyayn, Z.-Q. Luo, and C.~He, ``An iteratively weighted {MMSE} approach to distributed sum-utility maximization for a {MIMO} interfering broadcast channel,'' \emph{IEEE Trans. Signal Process.}, vol.~59, no.~9, pp. 4331--4340, Sep. 2011.

\bibitem{10741192}
Z.~Xiao, X.~Pi, L.~Zhu, X.-G. Xia, and R.~Zhang, ``Multiuser communications with movable-antenna base station: Joint antenna positioning, receive combining, and power control,'' \emph{IEEE Trans. Wireless Commun.}, vol.~23, no.~12, pp. 19\,744--19\,759, Dec. 2024.

\bibitem{10818453}
J.~Ding, Z.~Zhou, and B.~Jiao, ``Movable antenna-aided secure full-duplex multi-user communications,'' \emph{IEEE Trans. Wireless Commun.}, vol.~24, no.~3, pp. 2389--2403, Mar. 2025.

\bibitem{5447068}
Z.-q. Luo, W.-k. Ma, A.~M.-c. So, Y.~Ye, and S.~Zhang, ``Semidefinite relaxation of quadratic optimization problems,'' \emph{IEEE Signal Processing Magazine}, vol.~27, no.~3, pp. 20--34, May 2010.

\bibitem{Feng2024Weighted}
B.~Feng, Y.~Wu, X.-G. Xia, and C.~Xiao, ``Weighted sum-rate maximization for movable antenna-enhanced wireless networks,'' \emph{IEEE Wireless Commun. Lett.}, vol.~13, no.~6, pp. 1770--1774, Jun. 2024.

\bibitem{Ma2024MIMO}
W.~Ma, L.~Zhu, and R.~Zhang, ``{MIMO} capacity characterization for movable antenna systems,'' \emph{IEEE Trans. Wireless Commun.}, vol.~23, no.~4, pp. 3392--3407, Apr. 2024.

\bibitem{sun2016majorization}
Y.~Sun, P.~Babu, and D.~P. Palomar, ``Majorization-minimization algorithms in signal processing, communications, and machine learning,'' \emph{IEEE Trans. Signal Process.}, vol.~65, no.~3, pp. 794--816, Feb. 2016.

\end{thebibliography}
\end{document}